\documentclass[12pt,preprint]{aastex}

\shorttitle{Multi-wavelength properties of S5 0716+714}

\shortauthors{Liao et al.}

\begin{document}

\title{Multi-wavelength variability properties of {\it Fermi} blazar S5 0716+714}

\author{N. H. Liao\altaffilmark{1,2,3}, J. M. Bai\altaffilmark{1,2}, H. T. Liu\altaffilmark{1,2}, S. S. Weng\altaffilmark{4}, Liang Chen\altaffilmark{5} and F. Li\altaffilmark{1,2}
\thanks{E-mail:liaonh@ynao.ac.cn (NHL); baijinming@ynao.ac.cn (JMB)}}

\altaffiltext{1}{Yunnan Observatories, Chinese Academy of Sciences, Kunming, Yunnan 650011,
China}

\altaffiltext{2} {Key Laboratory for the Structure and Evolution
of Celestial Objects, Chinese Academy of Sciences, \\ Kunming,
Yunnan 650011, China}

\altaffiltext{3} {University of Chinese Academy of Sciences,
Beijing 100049, China}

\altaffiltext{4} {Department of Physics, Xiangtan University, Xiangtan 411105, China}

\altaffiltext{5} {Key Laboratory for Research in Galaxies and
Cosmology, Shanghai Astronomical Observatory, Chinese Academy of
Sciences,80 Nandan Road, Shanghai 200030, China}

\begin{abstract}

S5 0716+714 is a typical BL Lacertae object. In this paper we present
the analysis and results of long term simultaneous observations in
the radio, near-infrared, optical, X-ray and $\gamma$-ray bands, together
with our own photometric observations for this source. The light
curves show that the variability amplitudes in $\gamma$-ray and
optical bands are larger than those in the hard X-ray and radio
bands and that the spectral energy distribution (SED) peaks move
to shorter wavelengths when the source becomes brighter, which
are similar to other blazars, i.e., more
variable at wavelengths shorter than the SED peak frequencies.
Analysis shows that the characteristic variability timescales in
the 14.5 GHz, the optical, the X-ray, and the $\gamma$-ray bands
are comparable to each other. The variations of the hard X-ray and
14.5~GHz emissions are correlated with zero-lag, so are the V band
and $\gamma$-ray variations, which are consistent with the
leptonic models. Coincidences of $\gamma$-ray and optical flares
with a dramatic change of the optical polarization are detected.
Hadronic models do not have the same nature explanation for these
observations as the leptonic models. A strong optical flare
correlating a $\gamma$-ray flare whose peak flux is lower than the
average flux is detected. Leptonic model can explain this
variability phenomenon through simultaneous SED modeling. Different leptonic models are
distinguished by average SED modeling. The synchrotron plus
synchrotron self-Compton (SSC) model is ruled out due to the
extreme input parameters. Scattering of external seed photons,
such as the hot dust or broad line region emission, and the SSC
process are probably both needed to explain the $\gamma$-ray
emission of S5 0716+714.
\end{abstract}

\section{INTRODUCTION}

Blazars, including Flat Spectrum Radio Quasars (FSRQs) and BL
Lacertae objects (BL Lac objects), are radio-loud Active Galactic
Nuclei (AGNs, Urry \& Padovani 1995 and references therein). They
are characterized by the luminous, rapidly variable and polarized
non-thermal continuum emissions, extending from radio to
$\gamma$-ray (GeV and TeV) energies, which are widely accepted to
be produced in the relativistic jets oriented close to the line of
sight (Blandford \& Rees 1978; Ulrich et al. 1997). Their spectral
energy distributions (SEDs) have a universal two-bump structure in
log$\nu F_{\nu}$ - log$\nu$ representation, indicating two
different origins. The first bump is almost certain to be caused
by synchrotron emission of relativistic electrons as evidenced by
its high polarization from radio through optical wavelengths,
peaking at UV/X-rays in high-frequency peaked BL Lac objects
(HBLs) (Padovani \& Giommi 1995) and at IR/optical wavelengths in low-frequency peaked BL
Lac objects (LBLs) and FSRQs. The second
bump peaks at $\gamma$-rays whose origin is less well understood.

For the second bump, there are two types of emission models that both can well fit the observational SEDs. In
leptonic scenarios, the $\gamma$-rays are interpreted as inverse
Compton (IC) scattering of soft photons by relativistic electrons
which produce the first bump through synchrotron process. The soft
photons could be synchrotron photons [the Synchrotron Self-Compton
(SSC) model (Maraschi, Ghisellini \& Celotti 1992)] or photons
outside the jet [the External Radiation Compton (ERC) models: the
accretion disk radiation (Dermer \& Schlickeiser 1993); UV
emission from broad line region (BLR) (Sikora et al. 1994);
infrared emission from a dust torus (B{\l}a{\.z}ejowski et al. 2000)]. In
hadronic scenarios, if relativistic protons in a strongly
magnetized environment are sufficiently accelerated, the
particle-photon interaction processes, the synchrotron radiation
of protons and the proton-proton interaction processes must be
taken into account (M{\"u}cke $\&$ Protheroe 2001; Aharonian 2000;
Beall \& Bednarek 1999).

Multiwavelength variability study provides a way to test these
emission models which make different predictions for the relative
flare amplitudes and the time lags (e.g. Sambruna 2007; Marscher
et al. 2008). In the leptonic models, since the same population of
electrons is responsible for emitting both spectral components,
correlated variations of fluxes at the low and high energy peaks
with no lags are expected. While in the hadronic models, no
necessary correlations between the spectral components are
expected. Correlated variations across SEDs observed in some
blazars seem to be consistent with the leptonic models, but the
so-called ``orphan flares" observed in some TeV blazars prefer the
hadronic scenarios (e.g. Bonning et al. 2012; B{\"o}ttcher 2005).
Long term optical variability monitoring program for blazars has
been performed at Yunnan Observatories (e.g. Bai et al. 1998).
After the successful launch of the {\it Fermi} $\gamma$-ray
telescope, correlation between optical and $\gamma$-ray variations
has been focused. Because it is a bright typical BL Lac object and
sometimes peaks its synchrotron emission in the optical band, S5
0716+714 has been an important target in our monitoring list.

In this paper, we present our photometric observations of S5
0716+714, together with simultaneous multi-wavelength observations
from radio to GeV $\gamma$-rays acquired from public data archives
and literatures, to investigate its variability properties and
distinguish different radiation models. The outline of this paper
is as follows: in next section we give the review of the
historical information on S5 0716+714; the observation and data
reduction are shown in section 3; in section 4 we present the
variability properties and their implications; in section 5 we
report the SED modeling; in section 6 we present the discussions
and conclusions are in section 7. Throughout this paper, we refer
to a spectral index $\alpha$ as the energy index such that $
F_{\nu}\propto\nu^{-\alpha}$, corresponding to a photon index
$\Gamma_{ph} = \alpha+1$. We assume a $\Lambda$CDM cosmology with
$\rm H_{0}$ = 70 km $\rm s^{-1} Mpc^{-1}$, $\Omega_{m}$ = 0.27,
$\Omega_{\Lambda}$ = 0.73 (Komatsu et al. 2011).

\section{HISTORICAL OBSERVATIONS OF S5 0716+714}

S5 0716+714 is one of the brightest and most variable blazars. It
was discovered in the Bonn-NRAO radio survey (K{\"u}hr et al 1981)
and identified as a BL Lac object (Biermann et al.1981). The
source was classified as a LBL (Padovani \& Giommi 1995) or an
intermediate synchrotron peaked (ISP; Abdo et al. 2010a) blazar.
Nilsson et al. (2008) derived a redshift of z = 0.31 $\pm$ 0.08
(1$\sigma$ error) by the detection of its host galaxy. The
redshift was confirmed by the direct constraint from intervening
absorption lines, $0.23<z\lesssim0.37$ (99.7\% confidence)
(Danforth et al.2013).

S5 0716+714 is a famous intraday variability (IDV) source (Wagner
\& Witzel 1995). The IDV in radio band is possibly
source-intrinsic, rather than solely interpreted by interstellar
scintillation (Wagner et al. 1996; Kraus et al. 2003). Fuhrmann et
al. (2008) found that the flux densities from cm band to sub-mm
band were correlated and the radio spectra were inverted at about
86~GHz. The atypically fast apparent speeds of jet components were
from 4.5c to 16.1c ($z=0.3$), based on the last ten years Very
Long Baseline Interferometry data of S5 0716+714 (Bach et al.
2005).

No emission lines were detected in the IR, optical and
UV spectra of S5 0716+714 (Chen \& Shan 2011; Shaw et al 2009;
Danforth et al. 2013). The source is well known for almost
uninterrupted variability at multiple timescales. Extreme optical variations with maximum
variability rates of $\sim0.4$ mag $\rm hr^{-1}$ were
detected (Chandra et al. 2011; Danforth et al. 2013).
Not only as a famous IDV source, but S5 0716+714 is a source possessing variations on timescale of years
in the radio and optical bands, which may associate with changes
in the structure and/or direction of the inner jet (e.g. Raiteri et al. 2003; Nesci et al. 2005).
Strong optical polarized emissions, over 20\%, were observed (Takalo
et al. 1994; Ikejiri et al. 2011). Larionov et al. (2008) reported
a coincidence of a huge optical outburst with a $360\degr$
rotation of optical polarization angle (PA) in 2008 April.

Rapid X-ray flux variation with doubling timescale of $\sim$
7000 s was observed (Cappi et al. 1994). On timescale of hours, large and rapid variations were only
detected in the soft X-rays. Concave shape X-ray
spectra of S5 0716+714 were observed, with break energies of
$\sim$3~keV, exhibiting a steeper-when-brighter behavior.
It is considered that the soft X-rays are contributed by the
synchrotron emission from the high energy tail electrons while the
hard X-rays are from the IC process by the low energy electrons
(Giommi et al. 1999; Tagliaferri et al. 2003; Zhang 2010).

The source exhibits strong emission in $\gamma$-ray band. It was
detected by the EGRET on the Compton $\gamma$-ray
Observatory (CGRO) (Hartman et al. 1999; von Montigny et al.
1995). EGRET observation showed considerable $\gamma$-ray flux
variability with the stable spectra index within the statistical
uncertainty (Lin et al. 1995). In 2007 September, the
$\gamma$-ray flux detected by AGILE (Tavani et al. 2008)
exhibited an increase in flux by a factor of four in three days
(Chen et al. 2008). Since the launch of {\it Fermi}, S5 0716+714
had been in the list of the LAT (Atwood et al. 2009) Bright AGN Sample (LBAS;
Abdo et al. 2009), the First LAT AGN Catalog (Abdo et al. 2010d)
and the Second LAT AGN Catalog (2LAC; Ackermann et al. 2011). In
the investigation of the $\gamma$-ray energy spectra of LBAS
sources, a single power-law gave an acceptance to S5 0716+714 (Abdo
et al. 2010c). In 2LAC, a strongly caved spectrum
was observed and modeled by LogParabola function (Ackermann et al.
2011). A Very High Energy (VHE) $\gamma$-ray excess of S5 0716+714
was detected by MAGIC (Albert et al. 2008) and a possible
correlation between the VHE $\gamma$-ray and optical emissions was
suggested (Anderhub et al. 2009).

S5 0716+714 is active at all electromagnetic bands and several
papers have attempted to get insight on its property by
the SED modeling (e.g. Ghisellini et al. 2010; Zhang et al. 2012).
The inhomogeneous model could better explain the
ultra-fast X-ray variation than the homogeneous model (Ghisellini et al. 1997).
Tagliaferri et al. (2003) found that the single SSC model could not explain
the flat EGRET $\gamma$-ray spectrum and scattering of the
external soft photons probably needed to be considered (e.g. BLR;
accretion flow). Strongly variable optical and soft X-ray fluxes
with nearly constant $\gamma$-ray flux at 2007 October were
detected by GASP-WEBT-AGILE observations. This multi-wavelength
evolution was explained by the presence of two SSC components, one
of which is constant with the highly variable second one over the
entire observing period (Giommi et al. 2008).

After we submitted this paper, several papers aiming on the
variability of S5 0716+714 were published (Rani et al 2013a,b;
Larionov et al. 2013). Rani et al (2013a) also focuses on the
multi-wavelength variability from radio to $\gamma$-rays.
Rani et al (2013b) specially aims on the GeV $\gamma$-ray variability.
Larionov et al. (2013) analyzes the multi-wavelength outburst at
2011 October through the $\gamma$-ray, optical photometric and
polarimetric and Very Long Baseline Array (VLBA) observations.
\section{OBSERVATION AND DATA REDUCTION}
\subsection{Photometric Observation and Data Reduction}
The variability of S5 0716+714 was photometrically monitored in
the optical bands at Yunnan Observatories, making use of the 2.4 m
telescope\footnote{http://www.gmg.org.cn} and the 1.02 m
telescope\footnote{http://www1.ynao.ac.cn/$\sim$omt/}. The 2.4 m
telescope, started to work in May 2008, is located at $Lijiang$
observatory of Yunnan Observatories, where the longitude is
100\arcdeg01\arcmin51\arcsec E and the latitude is
26\arcdeg42\arcmin32\arcsec N, with a altitude of 3193 m. There
are two photometric terminals. PI VersArry 1300B CCD camera with
1340$\times$1300 pixels covers a field of view
4\arcmin48\arcsec$\times$4\arcmin40\arcsec at the Cassegrain
focus. The readout noise and gain are 6.05 electrons and 1.1
electrons$/$ADU, respectively. The Yunnan Faint Object
Spectrograph and Camera (YFOSC) has a field of view of about
10\arcmin$\times10$\arcmin and 2k$\times$2k pixels for photometric
observation. Each pixel corresponds to 0.283 arcsec of the sky.
The readout noise and gain of YFOSC CCD are 7.5 electrons and 0.33
electrons$/$ADU, respectively. The 1.02 m telescope is within the
headquarter of Yunnan Observatories, mainly used for photometry
with standard Johnson $UBV$ and Cousins $RI$ filters. An Andor CCD
camera with 2048$\times$2048 pixels had been installed at its
Cassegrain focus since May 2008. The readout noise and gain are
7.8 electrons and 1.1 electrons$/$ADU, respectively.

Sky flat field at dusk and dawn in good weather conditions and
bias frames were taken at every observing night. Because of the
negligible dark-current dark frame was skipped. Different exposure
times were applied for various seeing and weather conditions. All
frames were processed using bias and flat-field corrections by the
task \texttt{CCDRED} package of the \texttt{IRAF} software, while
the photometry was performed by the \texttt{APPHOT} package.
Magnitude of the source was calculated by differential photometry
with calibration stars in the image frame (Villata et al. 1998;
Ghisellini et al. 1997). Observing uncertainty of every night was
the root-mean-square (RMS) error of differential magnitude between
two calibration stars. At least one of them must be fainter than
or as bright as the source (Bai et al. 1999).
\begin{equation}
\sigma=\sqrt{\frac{\sum{\delta_i^2}}{N-1}},  i=1,2,\ldots N,
\end{equation}
where
\begin{math}
\delta=(m_2-m_1)_i-\overline{m_2-m_1}, ~\overline{m_2-m_1}
\end{math}
is the mean differential magnitude, $N$ is the number of data
points in the night.  The correction for the interstellar
extinction and the color excess were adopted according to Schlegel
et al. (1998). Optical photometric data were converted from
magnitude system to flux in Jansky (Bessell 2005). The results are
plotted in Figure 1a.

\subsection{Complementary Optical and other observations from the literatures}

Due to the observing time allocation, our photometric data
are sparse. In order to match the data-sampling of light curves at
X-ray and $\gamma$-ray energies, we collected V band photometric data of S5
0716+714 observed at the 1.5 m Kanata telescope of
Higashi-Hiroshima observatory in
Japan\footnote{http://hasc.hiroshima-u.ac.jp/telescope/kanatatel-e.html}
(Ikejiri et al. 2011) and the 2.3 m Bok telescope and 1.54 m
Kuiper telescope at Steward observatory of university of
Arizona (Smith et al. 2009) which were also adopted in Rani et al. (2013a), and from two
published papers (Poon et al. 2010; Chandra et al. 2011). The V
band light curve (Figure 2c) thus contain 324 data points,
extending from 54613.0 to 55822.0 MJD.

As mentioned in the Instruction Section, blazars are
characterized by their polarized continuum emissions. High linear
polarization from radio through optical wavelengths are common in
blazars, and usually the fractional variations in polarized flux
density are substantially larger than those in total flux density,
suggesting that polarization observation is an effective tool for
investigating emission process in blazar jets. Variability of
polarization in S5 0716+714 was monitored using the Kanata 1.5-m
telescope of Higashi-Hiroshima observatory from 2008 to 2009
(Ikejiri et al. 2009), and at the 2.3 m Bok telescope and 1.54 m
Kuiper telescope of Steward Observatory of the University of
Arizona since 2008 (Smith et al. 2009). The optical polarization,
flux, and spectral data of Steward Observatory are publicly
presented at their website\footnote{http://james.as.arizona.edu/$\sim$psmith/{\it Fermi}},
and those of Higashi-Hiroshima observatory have been published
(Ikejiri et al. 2011). These polarization
data were collected. The light curves of the V band PA and
polarization degree (PD) data are presented in Figure 2d
and 2e, respectively.

The synchrotron peak of S5 0716+714 shifts from time to
time in the near-infrared (NIR) and optical bands during different luminosity
states. The J and K bands photometric data of S5 0716+714 observed
at Higashi-Hiroshima observatory (Ikejiri et al. 2011) were also
included in this work. The K-band light curve is from 54613.0 to
54839.2 MJD, and J band light curve is from 54613.0 to 55228.0
MJD. The simultaneous light curves of the V, J, and K bands
obtained by the Kanata telescope are presented in Figure 1b.

The radio data were taken from the observations of 26 m
paraboloid of the University of Michigan Radio Astronomy
Observatory (UMRAO)\footnote{https://dept.astro.lsa.umich.edu/datasets/umrao.php}
(Aller et al. 1985, 1999). The light curves at 4.8~GHz,
8~GHz, and 14.5 GHz are from 54685.6 to 55917.2 MJD (Figure 2f).
These UMRAO multi-bands data from 54686 to 55600 MJD have been adopted in Rani et al. (2013a).

\subsection{X-ray Data Deduction}

X-ray data from Proportional Counter Array (PCA) in RXTE and X-ray
Telescope (XRT) in {\it Swift} are accessible. The PCA
observations are more continual than the XRT observations,
especially for the time range from 2009 to 2010. X-ray light curve
was extracted from the PCA data. S5 0716+714 is too faint to
obtain a detailed X-ray spectrum from PCA data whose energy range
is 2.6-50 keV which is beyond the spectral concave point of about
3 keV. X-ray spectra were extracted from XRT data with energy
range of 0.3-10 keV. The X-ray data were reduced by the FTOOLS
software package version 6.9.
\subsubsection{RXTE/PCA}

We analyzed RXTE observations taken between 2009 February 7 and
2010 December 28. We downloaded the Standard2 data from all layers of PCU2, that
operated during all the observations. The data were filtered
with the standard criteria: the Earth-limb elevation angle larger
than $10\degr$ and the spacecraft pointing offset less than
$0.02\degr$. The background files were created by using the
program \texttt{pcabackest} and the latest faint source background
model since the source intensity $<$ 40 counts/s/PCU (Weng \&
Zhang 2011). We applied the power law model to fit the RXTE/PCA
spectra over the energy range of 2.6-50.0 keV. An interstellar
absorption component with the neutral hydrogen column density
fixed to the Galactic value ($3.05\times10^{20}$ ${\rm cm^{-2}}$,
Murphy et al. 1996) was also included. The unabsorbed flux and its
error were also calculated in the 2.6-50.0 keV with the
\texttt{cflux} in XSPEC. The ten days bin X-ray light curve is represented
in Figure 2b.

\subsubsection{{\it Swift}/XRT}
The initial event cleaning was performed using the \texttt{xrtpipeline} script,
with the standard quality cuts (Weng \& Zhang 2011). The source spectra were
extracted with \texttt{xselect}, from circles with radius of 20
pixels centered at the nominal position of S5 0716+714, while the
background spectra were taken from a annulus regions with radius
30 and 60 pixels. If data were suffered from pileup, the annular
regions were used to describe the source, and the excluded region
radius depended on the current count rate
\footnote{http://www.{\it
Swift}.ac.uk/analysis/xrt/pileup.php}. We also produced the
ancillary response file with \texttt{xrtmkarf} to facilitate
subsequent spectral analysis. The response files (v013) were taken
from the CALDB database. Finally, the spectra were grouped to
require at least 20 counts per bin to ensure valid result using
$\chi^2$ statical analysis. The XRT spectra are exhibited in Table 1.

\subsection{{\it Fermi}/LAT Data Reduction}

The \texttt{Pass 7} $\gamma$-ray data were downloaded from LAT
data server, with time range from 4th August 2008 to 22th November
2011. The LAT event photons from 0.1 to 300~GeV were selected. The
LAT data analysis was performed with instrument response functions
of P7SOURCE\_V6, using the updated standard \texttt{ScienceTools}
software package version \texttt{v9r23p1}. For the LAT background
files, we used \texttt{gal$\_$2yearp7v6$\_$v0.fits} as the
galactic diffuse model and \texttt{iso$\_$p7v6source.txt} for the
isotropic spectral template\footnote{http://{\it
Fermi}.gsfc.nasa.gov/ssc/data/access/lat/BackgroundModels.html}.
For data preparation, \texttt{evclass=2} was adopted for
\texttt{gtselect}. The maximum zenith angle was set to $100\degr$.
We used unbinned likelihood algorithm (Mattox et al. 1996)
implemented in the \texttt{gtlike} task to extract the flux and
spectra. LogParabola function was used as the spectral model. All
sources from the Second {\it Fermi}/LAT catalog (2FGL, Nolan et
al. 2012) within $15\degr$ of the source position were included.
The flux and spectral parameters of sources within $10\degr$
``region of interest" (ROI) were set free, while parameters of sources
that fell outside the ROI were freezed at the 2FGL values.

For light curves, fit of each bin was scrutinized to make sure
that the fit quality is satisfactory and there is no background
source with negative TS value and exotic parameter. In a few low
TS cases, when the LogParabola function was not applicable, single
power-law model was used instead. There are only three fits with
TS values lower than 25 in the $\gamma$-ray light curve analysis
and the lowest TS value is 17 ($\geq 4\sigma$). We did not set
them as upper limits. The ten days bin $\gamma$-ray light curve is
shown in Figure 2a. The python script named
\texttt{SED$\_$scripts$\_$v13.1} from $Fermi$ User Contributions
was used to obtain $\gamma$-ray spectra. In the spectral analysis,
when the TS value was lower than 25, flux was replaced by
2$\sigma$ upper limit. All errors reported in the figures or
quoted in the text for $\gamma$-ray are 1$\sigma$ statistical
errors. The estimated relative systematic uncertainties on the
flux and effective area, are set to $10\%$ at 100MeV, $5\%$ at
500MeV and $20\%$ at 10GeV (Abdo et al. 2010b).

\section{THE VARIABILITY PROPERTIES}
\subsection{Spectral and Flux Variability around SED Peaks}
\subsubsection{Spectral Variability in the NIR-Optical Bands}

The peak frequencies and their changes of blazar SED are important
to obtain the parameters of emission model. The SEDs from NIR to
optical of S5 0716+714 are obtained from simultaneous V, J, and K
bands photometric data around the maximum flux at 54754.3 MJD when
the flux variation is intense and the time covering of the
observations is good (see Figure 3). The data observed at 54748.3
MJD are abandoned for the large error and peculiarly low flux in K
band. The SEDs are characterized by the inversions around J band in
relatively low condition, which suggests that the synchrotron peak
is likely close to the J band at the observational frame. When the
source flares, SEDs become flat between J and V bands. The SED is
not inverted at 54752.3 MJD, which means the frequency of the
synchrotron peak is higher than V band. Similar
bluer-when-brighter behavior has been found by Villata et al.
(2008) from contemporaneous GASP-WEBT to {\it Swift}/UVOT data for S5
0716+714.
\subsubsection{Spectral Variability at $\gamma$-rays}
The fit for 40 months LAT data is accomplished and the result gives
\begin{equation}
\frac{dN}{dE}=(1.27  \pm 0.02)\times 10 ^{-10}(\frac{E}{\rm 428.66
MeV})^{-((2.03 \pm 0.02) + (0.03 \pm 0.007) log(\frac{E}{\rm
428.66 MeV}))},
\end{equation}
with the average flux of (22.8 $\pm$ 0.4)$\times 10^{-8}$ ph $\rm
cm^{-2} s^{-1}$ which is similar to the EGRET detection of (2.0
$\pm$ 0.4) $\times 10^{-7}$ ph $\rm cm^{-2} s^{-1}$ (Lin et al.
1995). The 40 months LAT data is fitted by 20 energy bins.
The average spectrum departs from power law over
99\% confidence tested by the spectral curvature index $\mathcal {C}$
(Abdo et al. 2010b). A broken power law also gives an acceptance to the spectrum,
including a flat component between 0.1 and 1~GeV ($\Gamma=2.033\pm0.004$)
and a descent one up to higher energy ($\Gamma=2.282\pm0.023$). The index
of the decent part is in accordance with the non-simultaneous TeV
deabsorbed photon index of 1.8$\pm$0.6 (Anderhub et al. 2009). The
highest energy photon event with the probability of 0.9998 by
\texttt{gtsrcprob} is 207.4~GeV. It is much higher than 70~GeV for
BL Lacertae during the first 18-month period (Abdo et al. 2011).

Simultaneous $\gamma$-ray spectra are used to study the
$\gamma$-ray spectral variability of the source. Spectra of five
strongest $\gamma$-ray flares which peak at 54807.7, 55107.7,
55627.7, 55757.7 and 55857.7 MJD, correspond to the flaring state.
Another spectrum is obtained using $\gamma$-ray data from 55147.7
to 55207.7 MJD when the flux is in the low state. In the energy
range from 0.1~GeV to 1~GeV, the flaring state spectra are flat or
even ascending, which are harder than the low state spectrum.
Spectral variability around the second SED bump is similar with it
around the first SED bump. The general trend that when flux raises
the peaks of the SED bumps become bluer is the classical behavior
of SED evolution of blazars (Ulrich et al. 1997). The low state
spectrum and a flaring state spectrum corresponding to the
strongest flare, together with the average spectrum, are shown in
Figure 4.

\subsubsection{The Shortest Variability Timescale at $\gamma$-rays}

S5 0716+714 underwent three strong $\gamma$-ray flares in 2011.
Their peak fluxes are almost three times as the average flux. The
highest daily flux is $(1.69\pm0.25)\times10^{-6}$ ph $\rm cm^{-2}
s^{-1}$ at 55854.2 MJD which is lower than the AGILE detection of
$(2.03\pm0.75)\times10^{-6}$ ph $\rm cm^{-2} s^{-1}$ (Chen et al.
2008). The most intense variation appears when the flux increases
from $(3.74\pm1.67)\times10^{-7}$ ph $\rm cm^{-2} s^{-1}$ at
55853.2 MJD to $(1.69\pm0.25)\times10^{-6}$ ph $\rm cm^{-2}
s^{-1}$ at 55854.2 MJD. The $\gamma$-ray flux varies roughly 4.5
times at the interday timescale, which is more violent than a flux
increase by a factor of four in three days (Chen et al. 2008). The
doubling time less than one day in this flare agrees on the
finding of Rani et al. (2013b). Such a rapid $\gamma$-ray
variation allows us to make a constraint on Doppler factor
avoiding the heavy absorption from $\gamma\gamma$ process
(Begelman et al. 2008). The doubling time is about 21 hours and
the highest energy bin in the average $\gamma$-ray spectrum with
$\rm TS\geq25$ centers at about 72~GeV. Using the equation in
Dondi $\&$ Ghisellini (1995), $\delta\geq7.5$. Rani et al. (2013b)
uses the highest photon of 207.4~GeV as the absorbed photon and
makes the constraint of $\delta\geq9.1$.

\subsection{Multi-wavelength Correlations}
\subsubsection{Correlations of Radio/X-ray and Optical/$\gamma$-ray Variations}

The PCA X-ray and 14.5 GHz light curves seem to be well
correlated. Three flares of 14.5 GHz band at 55186.2, 55305.9 and
55475.5 MJD respectively correspond to the X-ray flares at
55185.6, 55302.8 and 55464.5 MJD (see three dotted vertical lines
in Figure 2). The low states between these flares in both energy
bands are also corresponding. However, the outlines of 14.5 GHz
flares are probably broader than those in the X-rays. No obvious
correlation can be directly seen between other two radio bands and
the PCA X-ray light curves. The flaring behaviors seem to be
washed out at 4.8 and 8~GHz.

Searching the existence of the correlation between optical and
$\gamma$-ray variations had been performed for S5 0716+714 since
the CGRO era (Ghisellini et al. 1997). Although the optical data in
our work is limited, most optical flares have the corresponding
$\gamma$-ray ones. The optical flux raises quickly during the
ascent phase of the strong $\gamma$-ray flare at 55627.7 MJD (see
the violet vertical line in Figure 2). Even in the extreme low
state of $\gamma$-ray, there probably exists a $\gamma$-ray flare
corresponding to a strong optical flare. On the
other side, three strongest optical flares with nearly constant
optical peak fluxes, correspond to three $\gamma$-ray flares with
their $\gamma$-ray peak fluxes changing three times. A
$\gamma$-ray flare at 54807.7 MJD with the peak flux of
$(2.36\pm0.61)\times10^{-7}$ ph $\rm cm^{-2}s^{-1}$ just above the
average flux, corresponds to the strongest optical flare at
54804.3 MJD. One of the strongest $\gamma$-ray flare at 55107.7
MJD when the peak flux is $(4.41\pm0.45)\times10^{-7}$ ph $\rm
cm^{-2} s^{-1}$, correlates a strong optical flare at 55115.3 MJD.
In this case, the $\gamma$-ray peak flux is almost twice as the
average flux. During the long low state of $\gamma$-ray from
55132.7 to 55222.7 MJD, another $\gamma$-ray flare at 55187.7 MJD with the peak
flux of $(1.42\pm0.38)\times10^{-7}$ ph $\rm cm^{-2} s^{-1}$,
corresponds to a strong optical flare at 55185.9 MJD. Peak
flux of this $\gamma$-ray flare is nearly half of the average
flux.

\subsubsection{Coincidence of $\gamma$-ray Flux and Optical PA Variation}

We find the coincidence of a $\gamma$-ray flare with a dramatic
change of optical PA at 2011 March (see Figure 5).
Within $\sim$ 30 days, the $\gamma$-ray flux sharply increases
from the low state at 55597.7 MJD to the flare peak at 55627.7 MJD.
The optical PA sharply increases from $(19.9\pm0.1)\degr$ at
55595.3 MJD to $(146.6\pm0.1)\degr$ at 55599.2 MJD within 5 days and
then decreases to $(22.6\pm0.2)\degr$ at 55625.2 MJD within 27 days
(see two dashed vertical lines in Figure 5). Within the increasing
phase of PA, the rotation rate of PA is $25\degr$ per day. Within
the decreasing phase of PA, the rotation rate of PA is $4.6\degr$
per day. There is a
flare in the V-band flux corresponding to this sharp $\gamma$-ray
flare. At the increasing phase of this optical
flare, the optical PA has a dramatic change as in the sharp
$\gamma$-ray flare (see Figure 5).  Larionov et al. (2013) finds that a
$180\degr$ rotation of the position angle of the optical linear
polarization coincides with strong flares in $\gamma$-ray and
optical bands in 2011 October. Actually, these similar observational phenomena
have been found in two of the three strong $\gamma$-ray
flares in 2011. Larionov et al. (2008) reports
another coincidence of a huge optical outburst with a $360\degr$
rotation of optical PA in 2008 April
while simultaneous $\gamma$-ray observation is missing.
This phenomenon seems to be a common occurrence for the source.
A similar behavior has been found in 3C 279 by Abdo et al.
(2010e). It is suggested that the sharp $\gamma$-ray flare is
correlated with the dramatic change of optical polarization,
likely due to a single, coherent event, rather than a
superposition of multiple but causally unrelated, shorter duration
events. Co-spatiality of optical and $\gamma$-ray emission regions,
and a highly ordered jet magnetic field are indicated (Abdo et al. 2010e).

\subsubsection{Strong Optical and X-ray Activities at the $\gamma$-ray Low State}

An interesting variability phenomenon of S5 0716+714 is that
intense variations appear in the X-rays, the optical flux and PD
during the long low state of the $\gamma$-rays from 55132.7 to
55222.7 MJD.  The highest flux in the X-rays is
$(14.47\pm2.21)\times10^{-12}$ erg $\rm cm^{-2} s^{-1}$ at 55164.6
MJD, raising from $(5.27\pm1.94)\times10^{-12}$ erg $\rm cm^{-2}
s^{-1}$ at 55143.2 MJD. There is a strong secondary X-ray flare
whose peak flux is $(13.25\pm1.98)\times10^{-12}$ erg $\rm cm^{-2}
s^{-1}$ at 55185.6 MJD. The flux of X-ray becomes low for
$(5.00\pm1.68)\times10^{-12}$ erg $\rm cm^{-2} s^{-1}$ at 55213.0
MJD. The fluxes of the optical double peaks are both $27.4\pm0.3$
mJy at 55183.8 and 55185.9 MJD, with the maximum optical PD of
$(25.69\pm0.80)$\% at 55183.3 MJD. The optical flux rises from
$8.0\pm0.1$ mJy at 55149.9 MJD and drops to $6.2\pm0.03$ mJy at
55204.3 MJD. The lowest optical PD in this epoch is
$(1.69\pm0.69)$\% at 55173.7 MJD. There is a flare of 14.5 GHz
peaks at 55186.2 MJD. However, the duration of this radio flare is
longer than flares in other bands. A $\gamma$-ray flare with peak
flux of $(1.41\pm0.38)\times10^{-7}$ ph $\rm cm^{-2} s^{-1}$ at
55187.7 MJD corresponds to the optical, optical PD and the
secondary X-ray flares. The peak flux of this $\gamma$-ray flare
is lower than the averaged flux while the corresponding peak
fluxes in other bands maintain in high state. But it is nearly
twice as the flux at 55177.7 MJD of (0.63$\pm0.32)\times10^{-7}$
ph $\rm cm^{-2} s^{-1}$ and roughly 3.6 times as the flux at
55197.7 MJD of (0.39$\pm0.25)\times10^{-7}$ ph $\rm cm^{-2}
s^{-1}$. So the $\gamma$-ray variability amplitude is comparable
with other bands. During this $\gamma$-ray low state epoch, the TS
value for each bin is larger than 25, which makes
the variability amplitude reliable. The fluxes of X-rays and
$\gamma$-rays all vary nearly three times. The intrinsic
variability amplitudes of these bands could be higher due to their
ten days averaged fluxes. The flux of optical V band varies more
than four times. The most violent behavior is shown as over ten
times variation on the optical PD. The X-ray flare at 55143.2 MJD,
which leads the optical and $\gamma$-ray flares, is probably
orphan discussed in Rani et al. (2013a).

Similar work using data from {\it Swift} and AGILE satellites has been
performed (Giommi et al. 2008). The multi-wavelength observational
results are similar to ours. However, it is claimed that the highly variable
optical and soft X-ray fluxes accompany with a
constant $\gamma$-ray flux which is based on the AGILE
observation with low counting statistics. Corresponding
$\gamma$-ray flare may be missed.

\subsection{Statistic Analysis on Multi-wavelength Light Curves}
\subsubsection{Time Lags}

The discrete correlation function (DCF; Edelson \& Krolik
1988) is a technique in time series analysis for finding time lags
between different light curves utilizing a binning scheme to
approximate the missing data. The z-transformed discrete
correlation function (ZDCF; Alexander 1997) can estimate the
cross-correlation function in the case of non-uniformly sampled
light curves. The ZDCF is a binning type of method as an
improvement of the DCF technique, with a notable feature that the
data are binned by equal population rather than equal bin width
$\Delta \tau$ as in the DCF. These light curves at the radio,
infrared, optical, and X-ray bands are sampled sparsely and
unequally for S5 0716+714. Thus, as in the previous researches on
the unequally sampled light curves (Liu et al. 2008, 2011a, b),
time lags will be analyzed by the ZDCF. From these ZDCF profile bumps
closer to the zero-lag, the centroid time lags $\tau_{\rm{cent}}$
which are used as the estimation of time lag,
are computed using all points with correlation coefficients $r
\geq 0.8 r_{\rm{max}}$.

The calculated ZDCFs are presented in Figures 6 for the light
curves in Figures 1 and 2. Considering the bin sizes of the X-ray
and $\gamma$-ray light curves, 14.5GHz/X-ray and the optical V
band/$\gamma$-ray variations appear to be zero-lag. The J-band
light curve (Figure 1), which is well sampled with shorter time
range than the V band light curve, also seems to zero-lag the
$\gamma$-ray light curve. We also use the classic DCF method to
check our correlation results. In general, the correlation results
from ZDCF and DCF are in agreement. The 14.5GHz/X-ray, the V
band/$\gamma$-ray and X-ray/$\gamma$-ray variations are all
strongly correlated with over 99.9\% confidence level. Rani et al
(2013a) and Larionov et al. (2013) also suggest the optical V
band/$\gamma$-ray variations are correlated with zero-lag.
However, we find that the relationship between the 14.5~GHz and
the $\gamma$-ray emissions is likely not tight. The correlation coefficient
is relatively low and the lag is not stable for different bin
sizes. Rani et al (2013a) finds that the confidence levels of the
correlations of $\gamma$-ray/37~GHz and  $\gamma$-ray/230~GHz are
lower than 3$\sigma$.
\subsubsection{Fractional Variability Amplitude}
In order to estimate the total variability of each light curve,
we use the RMS fractional variability amplitude
$F_{\rm{var}}$ (e.g., Edelson et al. 2002; Vaughan et al. 2003).
The fractional variability amplitude $F_{\rm{var}}$ is defined as
\begin{equation}
F_{\rm{var}}=\sqrt{\frac{S^2-<\sigma^2_{\rm{err}}>}{<F>^2}},
\end{equation}
where $<F>$ is the mean flux for the N points in the light curve,
$S^2$ denotes the total variance of the light curve, and
$<\sigma^2_{\rm{err}}>$ denotes the measured mean square error of
the data points:
\begin{equation}
S^2=\frac{1}{N-1}\sum^{\rm{N}}_{\rm{i=1}}(F_{\rm{i}}-<F>)^2,
\end{equation}
\begin{equation}
<\sigma^2_{\rm{err}}>=\frac{1}{N}\sum^{\rm{N}}_{\rm{i=1}}\sigma^2_{\rm{err,i}}.
\end{equation}
The error on $F_{\rm{var}}$ is (Edelson et al. 2002)
\begin{equation}
\sigma_{F_{\rm{var}}}=\frac{1}{F_{\rm{var}}}\sqrt{\frac{1}{2N}}\frac{S^2}{<F>^2}.
\end{equation}
For light curves in Figure 2 and the J band light curve in
Figure 1, the calculated results of $F_{\rm{var}}$ are listed in
Table 2. For all the light curves in Figures 1 and 2, the
calculated $F_{\rm{var}}$ are presented in Figure 7. The
fractional variability amplitude violently varies with the
frequency. Firstly, $F_{\rm{var}}$ increases with increasing
frequency within the radio band. This trend is same as that found in
3C 273 (e.g., Soldi et al. 2008). Secondly, $F_{\rm{var}}$ increases
from the radio to J band, decreases to V band and X-rays, and then
increases to $\gamma$-ray band (see Figure 7). Variability in the $\gamma$-ray band
is the most violent. The sampling rates and time intervals of
light curves significantly influence $F_{\rm{var}}$, and this is
clearly shown in the light curves at the K, J, I, R, V, and B
bands (see Figure 7).

\subsubsection{Characteristic Variability Timescale}

The zero-crossing time of the autocorrelation function (ACF) of a
light curve was defined as a single characteristic variability
timescale (Giveon et al. 1999). It is a well-defined quantity and
used as a characteristic variability timescale (e.g., Alexander
1997; Giveon et al. 1999; Netzer et al. 1996). Comparison of widths of the ACF between different bands can
shed light on the relation between the mechanism and
location of emission in these wave bands (Chatterjee et al. 2012). Another function
used in variability studies to estimate the variability timescale
is the first-order structure function (SF; e.g., Trevese et al.
1994). There is a simple relation between the ACF and the SF (see
eq. [8] in Giveon et al. 1999). Therefore, only an ACF analysis is
performed on our light curves. The ACF is estimated by the ZDCF
(Alexander 1997). Following Giveon et al. (1999), a fifth-order
polynomial least-squares procedure is used to fit the ZDCF, and
this fifth-order polynomial fit is used to evaluate the
zero-crossing time.

Calculated ACFs are presented in Figure 8 for the light curves in
Figure 2. These characteristic variability timescales of S5
0716+714 at the 14.5 GHz, V, X-ray, and $\gamma$-ray bands are
comparable to each other, $\sim$ 60--90 days. We also check the
effect of the binning on our results. Changing of bin size can not
significantly influence our tendency. These
comparable characteristic variability timescales indicate that
these emission variations likely have the same origin. While the origin
of 14.5~GHz emission could be complicated. Since the 14.5~GHz emission
is well correlated with the hard X-rays and the light curve of 14.5~GHz seems to be
less rapidly variable than other three bands, the 14.5~GHz emission is probably consisted of two
different components. One component could be more relative with the high energy emissions and from the compact radiation region. The other component is from a extended region and less rapidly variable. The characteristic variability timescale of 14.5~GHz from our ACF analysis likely corresponds to the first component. Rani et al.
(2013b) uses the SF analysis to obtain the $\gamma$-ray
variability timescale of $\sim$75 days. Rani et al. (2013a) finds
that the variability timescales of radio bands from 15~GHz to
230~GHz are in the range of $\sim$ 60--90 days by the SF analysis.
The variability timescale of the V band light curve is suggested
to be $\sim$60 days by Lomb-Scargle periodogram analysis (Rani et
al. 2013a). Our ACF results agree on these similar variability
timescale studies for S5 0716+714.

\subsection{Implication of Multi-wavelength Variability}

Concave X-ray spectra (Table 1) with the break energies of
$\sim$3~keV based on our XRT data analysis agree with the results
from literatures. The hard X-ray component is considered to be
from the IC process by the low energy electrons in the leptonic
models (Giommi et al. 1999; Tagliaferri et al. 2003; Zhang 2010).
Our X-ray light curve is obtained from the PCA data with energy
range of 2.6-50~keV. The radio emission of blazars is widely
accepted from the synchrotron process by the low energy electrons.
If the radio and the hard X-ray emissions both come from the low
energy electrons, 14.5~GHz and the PCA X-ray light curves should
be correlated. Such a prediction is confirmed by our correlation
analysis, strongly supporting the leptonic models.

Correlation with a zero-lag between the optical V band and
$\gamma$-ray variations is also found, which agrees on similar
works (Rani et al 2013a; Larionov et al. 2013). 3C 66A, with the
source type similar to S5 0716+714, shows a clear correlation
between GeV $\gamma$-ray and optical R band with a time lag of
$\lesssim$ 5 days (Reyes et al. 2011). Well correlations between
optical/IR and $\gamma$-rays in six blazars have been found with
zero-lags for FSRQs (Bonning et al. 2012). There are $\gamma$-ray
and optical flares at about 55183 MJD when a flare of optical PD
coincides. The coincidence of $\gamma$-ray and optical flares with
a dramatic change of optical PA at 2011 March is found. Similar
behavior happens at 2011 October (Larionov et al. 2013). All the
above observations suggest a tight relationship between the
optical and the $\gamma$-ray emissions. These multi-wavelength
variation phenomena accord with the leptonic models.

In view of the global frequency range, the variability amplitudes and characteristic variability
times of S5 0716+714 from radio to $\gamma$-rays can be naturally explained by the leptonic models.
We find that the $F_{\rm{var}}$ is a function of frequency. The $F_{\rm{var}}$ increases
from the radio band to optical band, decreases to hard X-rays, and then
increases to $\gamma$-rays. In the leptonic models, the electrons which generate the optical
and $\gamma$-ray emissions have higher energies than those for radio
and hard X-ray emissions. The radiation cooling time of the high
energy electrons is shorter than it of the low energy electrons.
Through our ACF analysis, the characteristic variability
times of 14.5~GHz, V band, X-ray and $\gamma$-ray are comparable to
each other, which agrees on the variability timescales from SF
analysis (Rani et al. 2013a,b). It is indicated that variations
of these emissions likely have the same origin. These four bands
emission are suggested to be from the same population of electrons
in the leptonic models. Although the variation origin of blazars is
extremely complicated, the variation of the properties of the
electrons can be one of the most important reasons for variability
of emissions of these four energies.

``orphan" $\gamma$-ray flare is not found by comparing the optical light curve.
Considering the strong correlation with a zero lag between optical and $\gamma$-ray flux variations
and coincidences of $\gamma$-ray and optical polarization variations, the hadronic models do
not have the same nature explanation for these observations than the leptonic models.

\section{SED MODELING}

The SED modeling is a powerful tool to test different radiation
models for blazars. In this paper, a homogeneous one zone
synchrotron plus IC model is used to calculate the jet emission of
S5 0716+714. The broadband electromagnetic emission comes from a
compact homogeneous blob with relativistic speed having the radius
of $R$ embedded in the magnetic field. A broken power law spectrum
for particle distribution has been assumed,
\begin{equation}
N(\gamma )=\left\{ \begin{array}{ll}
                    K\gamma ^{-p_1}  &  \mbox{ $\gamma_{\rm min}\leq \gamma \leq \gamma_{br}$} \\
            K\gamma _{\rm br}^{p_2-p_1} \gamma ^{-p_2}  &  \mbox{ $\gamma _{\rm br}<\gamma\leq\gamma_{\rm max}$.}
           \end{array}
       \right.
\label{Ngamma}
\end{equation}
Such a broken power law distribution can be the result of the
balance from the particle cooling and escape in the blob. The
parameters of this model include, the radius $R$ of the blob, the
magnetic field strength $B$, electron break energy $\gamma_{\rm
br}$, the minimum and maximum energy $\gamma_{\rm min}$,
$\gamma_{\rm max}$, of the electrons, the normalization of the
particle number density $K$, and the indexes $p_{1,2}$ of the
broken power law particle distribution. The frequency and
luminosity can be transformed from the jet frame to observational
frame as: $\nu=\delta\nu'/(1+z)$ and $\nu
L_{\nu}=\delta^{4}\nu'L_{\nu'}'$, where the Doppler factor
$\delta=1/\left[\Gamma\left(1-\beta\cos\theta\right)\right]$. The
synchrotron self-absorption and the Klein-Nishina effect in the IC
scattering are properly considered in our calculations. The
detailed constraints of the SED modeling can be found in Tavecchio
et al. (1998) and Sikora et al. (2009).

We use the $\chi^{2}$-minimization method to obtain the best
fitting input parameters. We make special constraints on parameters of B and $\delta$.
The values of these parameters are varied in wide ranges to calculate the
corresponding values of $\chi^{2}$. Then, we obtain the
probability of the fit by $p\propto e^{-\chi^{2}/2}$. We plot the
contours of p in the B-$\delta$ plane and constrain the value of
the B and $\delta$ at 1$\sigma$ level. Detailed SED modeling
strategy can be found in Zhang et al. (2012).

\subsection{Average SED modeling}
Thanks to the long accumulative observation time, the average
$\gamma$-ray spectrum from 0.1 GeV to almost 100 GeV with
relatively small error brings much more information than
observations before, the EGRET and AGILE observations (Lin et al.
1995; Chen et al. 2008). Modeling the average SED
with different models are shown in Figure 9 and the input
parameters are listed in Table~3.

S5 0716+714 possesses violent variations across its broadband
electromagnetic radiation. It is a famous IDV source in
the radio and optical bands, doubling time of 7000 s at X-ray, and
$\gamma$-ray flux changing 4.5 times for two days, which are
strict constraints on the emission region, $R\leq$
c$t_{var}\delta (1+z)^{-1}$ (Begelman et al. 2008). The Doppler
factor can be constrained by rapid variations ($\gtrsim$ 5-15, Fuhrmann et al. 2008) and kinematic study ($\approx$ 20-30, Bach et al. 2005).
The strength of the magnetic field can be constrained by the inverted radio spectra demonstrated as the result of
synchrotron self-absorption ($\gtrsim$ 0.07-0.11 $\delta$G, Fuhrmann et al. 2008)

In Table~3, it can be seen that for pure SSC model which is
usually successful for TeV blazars, the input model parameters are
extreme. When the variability timescale is set to one day, the
Doppler factor value becomes extraordinarily large which conflicts
with the typical value from kinematic studies for blazars. The
Doppler factor is reduced to $39.6_{-3.1}^{+3.8}$ when variability
timescale is set to ten days. However, the variability timescale
of ten days disagrees on the observations of fast variability from
the radio to $\gamma$-rays. Meanwhile, the magnetic field B is
$0.006_{-0.002}^{+0.001}$ Gause which is not harmonious with the
typical value from similar SED fitting works before (eg.
Tagliaferri et al. 2003; Giommi et al. 2008). The large frequency
ratio of the SSC peak to synchrotron peak may bring out these
abnormal parameters. The pure SSC model is also not favorable due
to its failure for explaining the extreme fast optical variability
(Danforth et al. 2013).

Although no thermal components have been detected from
spectroscopic observations (Chen \& Shan 2011; Shaw et al 2009;
Danforth et al. 2013), scattering of weak external emissions could
possibly contribute the $\gamma$-ray emission of the source.
Because the synchrotron peak of S5 0716+714 locates at NIR/optical
band at the observation frame, the hot dust emission can be
heavily diluted by the strong nothermal continuum. A prominent
infrared excess indicative of dust emission with 1200 K in 4C
21.35 has been detected, which proves the hot dust emission can be
the main contribution to the external seed photons for IC process
(Malmrose et al. 2011). Both the assumed hot dust and BLR
emissions are individually considered in the average SED modeling.
The hot dust and BLR emissions are approximated as black body
emissions of the temperature of 1200 K and peaking at the
frequency of the Ly$\alpha$ line, respectively. For each fitting
process, the variability timescale is set to one day and the
photon energy density is set free.

The contours of p for the case of SSC+ERC model with the
hot dust photons are shown in Figure 10. The B
and $\delta$ distributions of p can be seen in Figure 11. The values of $\delta$ are
$24.5^{+0.6}_{-0.7}$ and $23.3^{+0.7}_{-0.9}$ corresponding to the hot
dust and BLR emissions as the external seed photons, respectively. These values
are more agreeable on the observations than the $\delta$ value
from the pure SSC model with variability timescale as one day.
The magnetic field intensity values are $0.24^{+0.02}_{-0.01}$ and
$0.28^{+0.02}_{-0.02}$ corresponding to the hot dust and BLR emissions
as the external seed photons, respectively. Although these values are still
lower than the constraint from radio spectra (Fuhrmann et al. 2008), they are more
reasonable than the magnetic field intensity from the pure SSC model. The SSC+ERC models
are more favorable avoiding the extreme input parameters and better
explaining the fast variability than the pure SSC model.

\subsection{Simultaneous SED modeling}

As mentioned in Section 4.2, three strong optical flares with
nearly constant optical peak fluxes correspond three $\gamma$-ray
flares with their $\gamma$-ray peak fluxes changing three times.
At 55187.7 MJD, peak flux of the $\gamma$-ray flare is even lower
than the average flux while peak fluxes of corresponding flares in
optical and X-ray bands maintain in high state. This seems to be
abnormal for the leptonic models.

Three simultaneous SEDs are obtained for optical and $\gamma$-ray
flaring state, together with another simultaneous SED for
multi-wavelength low state. For the first SED, the strongest
optical flare at 54804.3 MJD corresponding to a $\gamma$-ray flare
at 54807.7 MJD with peak flux just above the average flux is
focused. No simultaneous X-ray observation is found. The second
SED corresponds to one of the strongest $\gamma$-ray flare at
55107.7 MJD correlated with a strong optical flare at 55115.3 MJD.
X-ray constraint is the interpolation of the edges of the PCA
observation blank at 55091.0 and 55136.0 MJD. Violent X-ray and
optical activities have been detected during the long low
$\gamma$-ray period from 55132.7 to 55222.7 MJD. Two SEDs are
corresponding to the low and high states for highly variable X-ray
and optical fluxes. A individual $\gamma$-ray spectrum for the
flare at 55187.7 MJD can not be obtained due to shortage of enough
$\gamma$-ray photons. So both the high and low states SEDs share
the entire low state $\gamma$-ray spectrum. The optical peak flux
at 55185.9 MJD and X-ray spectrum from 55172.0 to 55187.0 MJD are
used for the flaring state SED. While the low optical flux at
55204.3 MJD and X-ray spectrum from 55193.0 to 55249.0 MJD are
adopted for the low state SED. The nearest UMRAO data are used as
upper limits.

We only adopt the SSC+ERC model with external seed photons from
hot dust emission. The typical variability timescale is set to one
day. The temperature and the photon density of hot dust emission
are fixed to be consistent with the input parameters of the
average SED modeling.

The input parameters from modeling the simultaneous SEDs are shown in Table
3 and the fit results are shown in Figure 12. Four simultaneous
SEDs are well described by the SSC+ERC model, which agrees with the result of similar
simultaneous SED modeling work from Rani et al. (2013a). The variability phenomenon
can be explained by the leptonic model. For S5 0716+714,
its V band emission is likely around the synchrotron
peak while the SSC emission peaks at about 10 MeV which is $\sim$1 order
of magnitude lower than the energy range of {\it Fermi}/LAT,
0.1-300 GeV. When the frequencies of IC peaks become lower, the GeV
$\gamma$-ray flux detected by LAT descends quickly. However, the V band
emission can not be influenced seriously by a little change of
synchrotron peak frequency because the SED slope is flat
around synchrotron peak. Flares at 55107.7 and 55187.7 MJD are
used to make a contrast. The input parameters at Table 3 are used.
Since $\nu_{syn}\propto\gamma_{\rm br}^{2}B\delta$ and
$\nu_{ssc}\simeq\gamma_{\rm br}^{2}\nu_{syn}$, the frequency of
synchrotron peak in former flare is roughly 1.2 times as the latter,
while for the frequencies of SSC peaks, it is about 2.8 times.
It accords with the observations that the optical peak fluxes are almost
same while the change of $\gamma$-ray peak fluxes is nearly three times.

The Doppler factors of three $\gamma$-ray flares
trend to have higher values comparing to the low state. It
corresponds with the VLBA observation that $\gamma$-ray flare
always accompanys with apparent super-luminal knot ejection. The
magnetic field intensities of these three $\gamma$-ray flares also
trend to have higher values than the low state. The highest
magnetic field intensity value is found in the flare at December
2009 when a extreme high PD of $(25.69\pm0.80)$\%
is detected. For the three $\gamma$-ray flares, the $\gamma$-ray
fluxes seem to be inversely proportional to magnetic field
intensity values and proportional to the the Doppler factor
values. However, the latter becomes marginal due to the fitting
errors. These two tendencies are similar to the finding for 3C
454.3 (Bonnoli et al. 2011) and 3C 279 (Zhang et al. 2013).

\section{DISCUSSIONS}

Due to the absence of emission lines in the optical spectra of S5
0716+714, its redshift has not been exactly determined. In the
above SED modeling, we took $z=0.31\pm0.08$ ($1\sigma$ error) from
Nilsson et al. (2008). Different radiation models are distinguished
by the input model parameters which depend on the redshift. We
attempt to discuss the possible influence caused by the
uncertainty of the redshift. We take the SSC+ERC model with the
hot dust emission as an example. 11 redshift points are chosen
evenly in range of the uncertainty of reshift (0.23,0.39). We
model the 11 SEDs with different redshifts independently. The B
and $\delta$ distributions of the reshift are shown in Figure 13.
The influence of $1\sigma$ uncertainty of reshift on B is
comparable with the $1\sigma$ fit error. However, it is larger
than the fit error on $\delta$. We also check the redshift
influence on the pure SSC model and the SSC+ERC model using BLR
emission. Including the influence of $1\sigma$ uncertainty of
reshift, the tendency that input parameters from the pure SSC
model are more extreme than the SSC+ERC models, is not changed.

To avoid the extreme input parameters from the pure SSC model, SSC+ERC models are used instead.
On the other side, the
assumption of the existence of the external emissions can not
conflict with any observations of S5 0716+714. The luminosity of the
presumed external emissions can not excess the nearby non-thermal jet
emission. As few information of the accretion system is known, the
characteristic radius scales of the dust torus $R_{dust}\sim1$ pc
and the BLR $R_{Ly\alpha}\sim0.01$ pc are used, then
\begin{equation}
L_{dust,Ly\alpha} \simeq 4\pi R_{dust,Ly\alpha}^{2}c U_{dust,Ly\alpha} ,
\end{equation}
where $U_{dust,Ly\alpha}$ is the energy density of the dust or
Ly$\alpha$ line emission, and $c$ is the speed of light. The energy
densities of the external field emissions are obtained from the
average SED modeling, $U_{dust}\simeq 1.1\times10^{-6}$ erg $\rm
cm^{-3}$ and $U_{Ly\alpha}\simeq 8.6\times10^{-6}$ erg $\rm
cm^{-3}$, at observational frame. So the $L_{dust}$ and $L_{Ly\alpha}$ are calculated as
$4.0\times10^{42}$ and $3.1\times10^{39}$ erg $\rm s^{-1}$, respectively. The non-thermal luminosity at the frequency
of the maximum hot dust emission is about $1.7\times10^{46}$ erg $\rm s^{-1}$ (Chen \& Shan 2011). The constraint of the Ly$\alpha$ line luminosity has been recently obtained, L(Ly$\alpha)\lesssim6.5\times10^{41}$ erg~$\rm s^{-1}$
(Danforth et al. 2013). So the luminosities of the external
emissions are lower than the nearby non-thermal luminosities, which makes our assumption harmonious.

Strong $\gamma$-ray emissions from S5 0716+714 have been detected by
MAGIC and {\it Fermi}. If the absorptions for $\gamma$-rays caused by the assumed external emissions are
significant, the $\gamma$-ray photons can not escape, which
conflicts with the $\gamma$-ray observations. The optical depth of
photon-photon absorption between the $\gamma$-rays and the
external emission can be simply calculated as (Dondi \& Ghisellini
1995),
\begin{equation}
\tau_{\gamma\gamma}(x^{'})=\frac{\sigma_{T}}{5}n^{'}(x^{'}_{t})x^{'}_{t}R^{'},
\end{equation}
where $\sigma_{T}$ is the scattering Tomson cross-section,
$n^{'}(x^{'}_{t})$ is the number density of target photon,
$x^{'}_{t}$ is the energy of target photon in  dimensionless units
and the $R^{'}$ is the absorption length. The doubling time of 21
hours is used to constrain the emission region. The absorption
length is considered the same as the distance between black hole
and the emission blob (Celotti et al. 1998). $R^{'}\simeq
R_{\gamma}=ct_{var}\delta^{2}(1+z)^{-1}$. No matter the external
seed photons are from the dust or Ly$\alpha$ line emission, the
absorption opacity $\tau_{\gamma\gamma}\leq10^{-6}\ll1$. The absorptions are
negligible. The assumptions of the external emission do not
conflict with the $\gamma$-ray observations.

Synchrotron plus SSC+ERC models can well describe the
broadband emission of S5 0716+714. Similar results are
found in serval LBLs/IBLs. For BL Lacertae, the prototype of the
BL Lac objects, shows that the SSC+ERC model using the BLR
emission as seed photons is better than a two-zone SSC model and
the pure SSC model is likely to be ruled out, acting like a FSRQ (Abdo
et al. 2011). The SSC+ERC model with external IR seed photons can
accord with the intraday variability for the 3C 66A (Reyes et al
2011), and get a more reasonable magnetic field parameter for
W Comae (Acciari et al. 2009). Actually, any evidence of the thermal
emission has not been found for S5 0716+714 and 3C 66A. However,
the external seed photons are probably necessary to explain their
$\gamma$-ray emissions. LBLs/IBLs, unlike the typical HBLs which
can usually be simply well explained by single synchrotron plus
SSC model (e.g. Bartoli et al.
2012), or the FSRQs whose the $\gamma$-ray emissions are
often dominated by the ERC process (e.g. Bonnoli et al. 2011), are the kind of sources that
both of the SSC and ERC processes may be indispensable for their
$\gamma$-ray emissions.

\section{CONCLUSIONS}
We present the results of the radio to $\gamma$-ray observations
of S5 0716+714, together with our photometric observations at
Yunnan Observatories. The variations of the 14.5 GHz and hard X-ray emissions are
correlated with zero-lag, which strongly supports the leptonic
models. A coincidence of $\gamma$-ray and V band flares with a
dramatic change of the optical PA at 2011 March is detected.
$\gamma$-ray and V band flares at 2009 December correspond to a
flare of optical PD. The V band and $\gamma$-ray flux
variations are also correlated with zero-lag, consistent with the
results of Rani et al. (2013a) and Larionov et al. (2013). A
tight relationship between the optical and $\gamma$-ray emissions
is suggested. The variability amplitudes in $\gamma$-ray and
optical bands are higher than those in the hard X-ray and radio
bands. The radiation cooling time of the high energy electrons
which radiate the optical and $\gamma$-ray emissions is much
shorter than it of the low energy electrons which produce the hard
X-ray and radio emissions. The characteristic timescales
of 14.5 GHz, optical, X-ray, and $\gamma$-ray bands from our ACF
analysis are comparable to each other, which are consistent with
results of the SF analysis of Rani et al. (2013a,b) that the
characteristic timescales of radio, optical and $\gamma$-ray are
$\sim$60-90 days. Variability of these bands are likely from the
same origin which could be the change of properties of the
radiating electrons. Hadronic models do not have the same nature
explanation for these observations than the leptonic models. By
comparing the optical light curve, ``orphan" $\gamma$-ray flare
which supports the hadronic models is not found. However, we find
a peculiar phenomenon that a strong optical flare correlates a
$\gamma$-ray flare whose peak flux is lower than the average flux.
Leptonic model can explain this variability phenomenon through
simultaneous SED modeling. Conclusively, the multi-wavelength
emissions of S5 0716+714 are likely generated from the
relativistic electrons.

Different leptonic models are distinguished by the average SED
modeling. The pure SSC model is ruled out due to the
extreme input parameters, according with result of Danforth et al.
(2013) that the pure SSC model is different to explain the extreme
fast optical variability. SSC+ERC model, whether the BLR or the
hot dust emission is used as the external emission for the IC
process, can well represent the SEDs and provide reasonable input
parameters. It agrees with Rani et al. (2013a) which suggests that
the SSC+EC model using BLR emission as the external photons
provides a satisfactory description of the broadband SEDs.
Including the influence of 1$\sigma$ uncertainty of the redshift of the source,
the tendency that SSC+ERC models are more favorable than the pure
SSC model has no change. The luminosities of the assumed external
emissions do not excess the luminosities of nearby jet continuous
emissions and the absorptions for $\gamma$-rays caused by the
assumed external emissions are negligible, which makes our
assumptions harmonious. Both SSC and ERC processes are probably
needed to explain the $\gamma$-ray emission of S5 0716+714.

\acknowledgements This research has made use of data obtained from
the High Energy Astrophysics Science Archive Research Center,
provided by NASA/Goddard Space Flight Center. We thank Makoto
Uemura who provides the published Kanata
optical/NIR flux and polarization data. Data from the
Steward Observatory spectropolarimetric monitoring project were
used. This program is supported by {\it Fermi} Guest Investigator
grants NNX08AW56G, NNX09AU10G, and NNX12AO93G. We thank Paul Smith who
provides the public V band flux and polarization data from Steward Observatory.
This research has made use of data from the
University of Michigan Radio Astronomy Observatory which has been
supported by the University of Michigan and by a series of grants
from the National Science Foundation, most recently AST-0607523.
We thank Margo Aller who provides the UMRAO multi-bands flux density data.
This work is financially supported by the 973 Program (Grant
2009CB824800), the National Natural Science Foundation of China
(NSFC; Grants 11133006, 11273052, 11233006, 11173043, 11133002 and 11103060)
and the Youth Innovation Promotion Association, CAS.

We appreciate the anonymous referee for the helpful suggestions
that led to a substantial improvement of this work.
We thank Jian Cheng Wang, Wei Cui, Jin Zhang for their fruitful
suggestions during different discussions. Yue Heng Xu is specially
thanked for improving the quality of English language. Shao Kun
li, Chuan Jun Wang, Yu Xin Xin and Xu Liang Fan are appreciated
for their observations at 1~m and 2.4~m telescopes of Yunnan Observatories. Yi Bo Wang is appreciated for his
suggestions of statistic analysis. {\it Fermi} help group,
especially for Robin Corbet and Jeremy S. Perkins, are appreciated
for their advices for the data analysis of {\it
Fermi}/LAT.

\begin{figure}
\centering
\includegraphics[width=3.5 in,angle=-90]{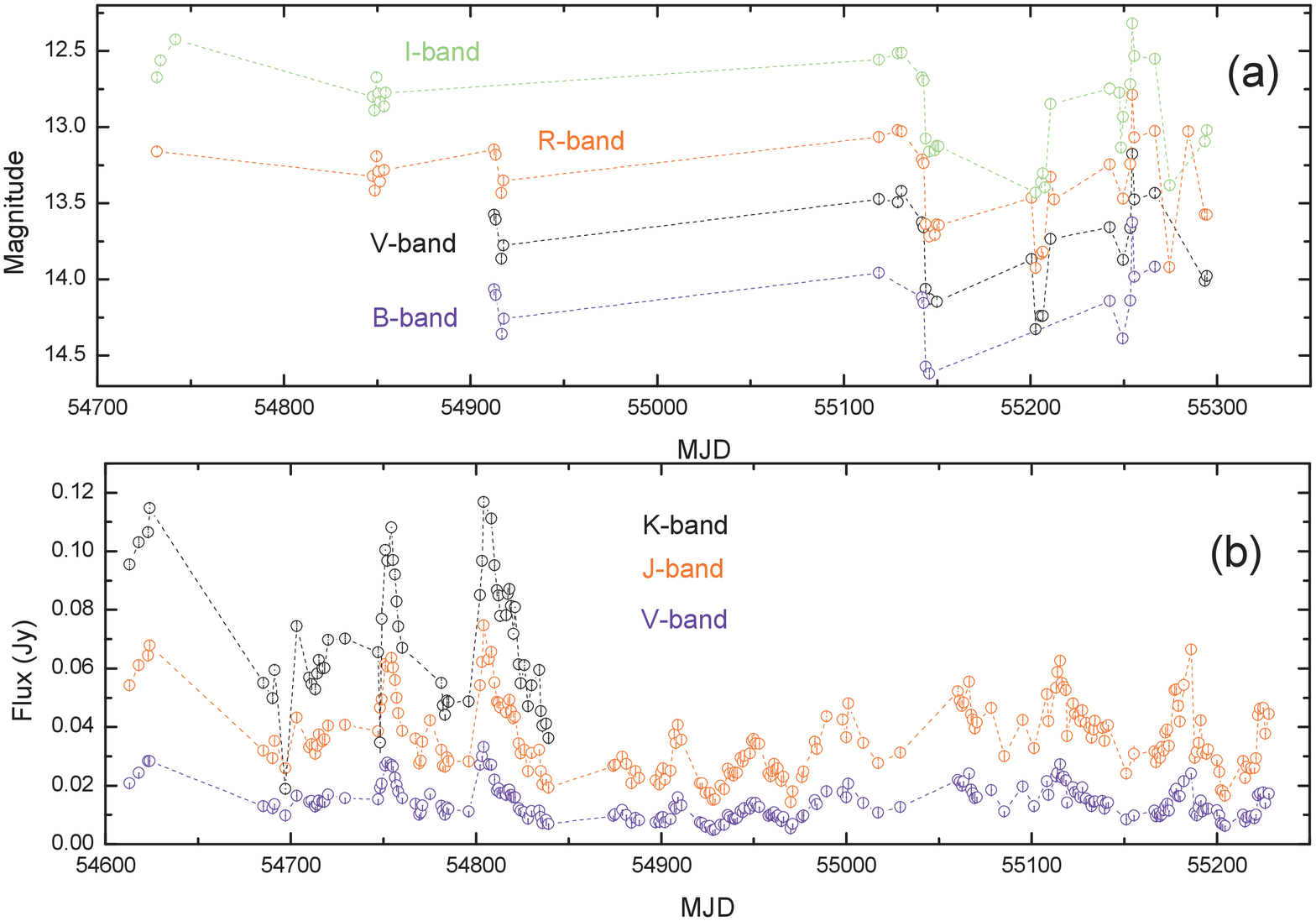}
 \caption{(a) Light curves at I, R, V, and B bands observed by Yunnan Observatories, (b) Light curves at K, J, and V bands of 0716+714 obtained from observations of Kanata telescope at Higashi-Hiroshima
observatory (Ikejiri et al. 2011).}
  \label{Figure 1.}
\end{figure}

\begin{figure}
\centering
\includegraphics[width=6.5 in]{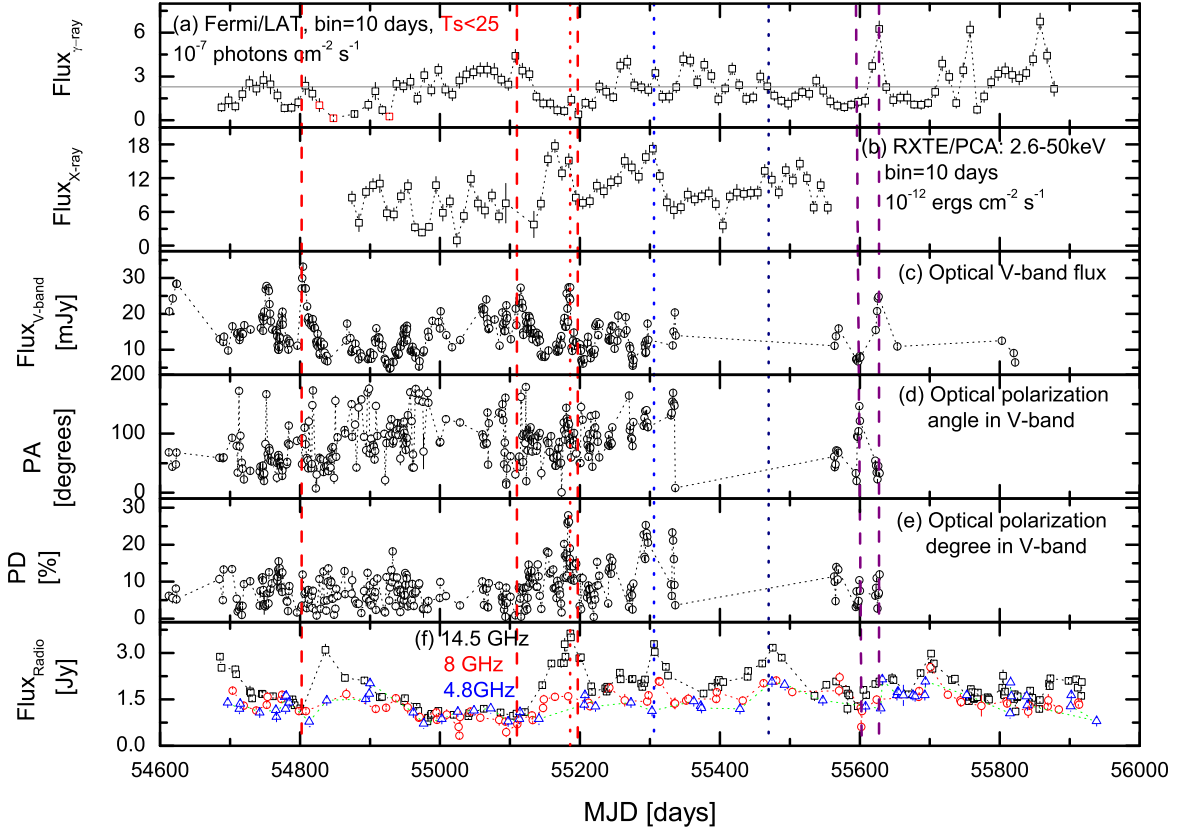}
 \caption{Light curves of radio (UMRAO), optical (detailed data origin can be seen at section 3), X-ray (RXTE/PCA), and $\gamma$-ray ({\it Fermi}/LAT) emissions. The horizontal line in plot (a) denotes the average of $\gamma$-ray flux. The four red vertical lines correspond to the four simultaneous SEDs in the text. The two dashed vertical violet lines at 55595 and 55625 MJD indicate the coincidence of a $\gamma$-ray flare with a dramatic change of optical polarization angle. This coincidence is zoomed in to Figure 5. The three dotted vertical lines indicate the peaks of the 14.5 GHz outbursts I, II, and III, which have three well correlated X-ray flares. }
  \label{Figure 2.}
\end{figure}

\begin{figure}
\centering
\includegraphics[width=3.5 in,angle=-90]{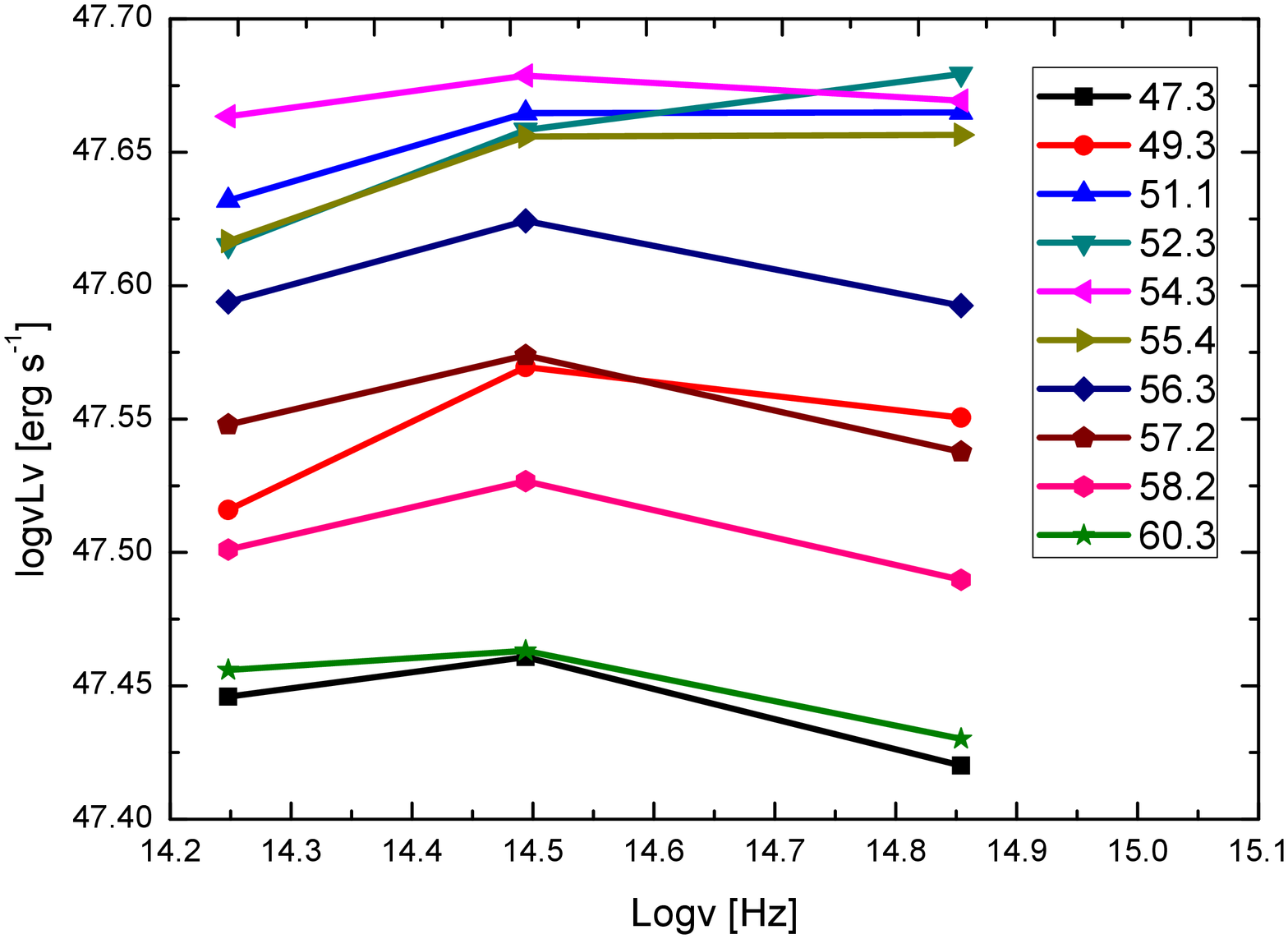}
 \caption{NIR-optical SEDs, constructed with simultaneous V, J, K
 data observed from Kanata telescope (Ikejiri et al. 2011). These data have been already corrected for the interstellar extinction and the color excess. The numbers represent the corresponding dates of the SEDs, MJD-54700.}
  \label{Figure 3.}
\end{figure}

\begin{figure}
\centering
\includegraphics[width=3.5 in,angle=-90]{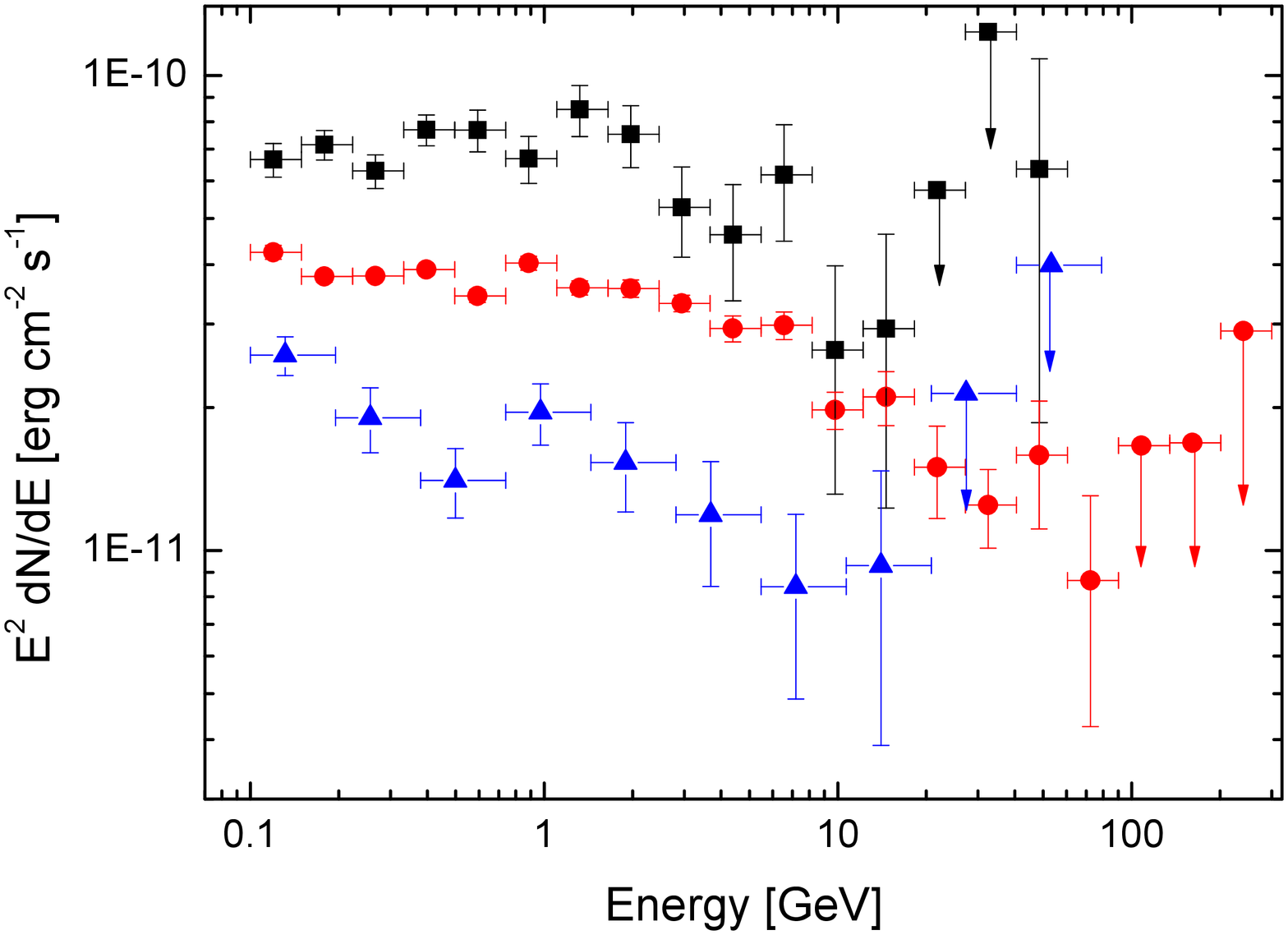}
 \caption{$\gamma$-ray spectra in different states. The black squares represent the spectrum corresponding to the strongest flare at 55857.7 MJD. The red circles represent the average spectrum. The blue triangles correspond to the low state spectrum.}
  \label{Figure 4.}
\end{figure}

\begin{figure}
\centering
\includegraphics[width=5.5 in]{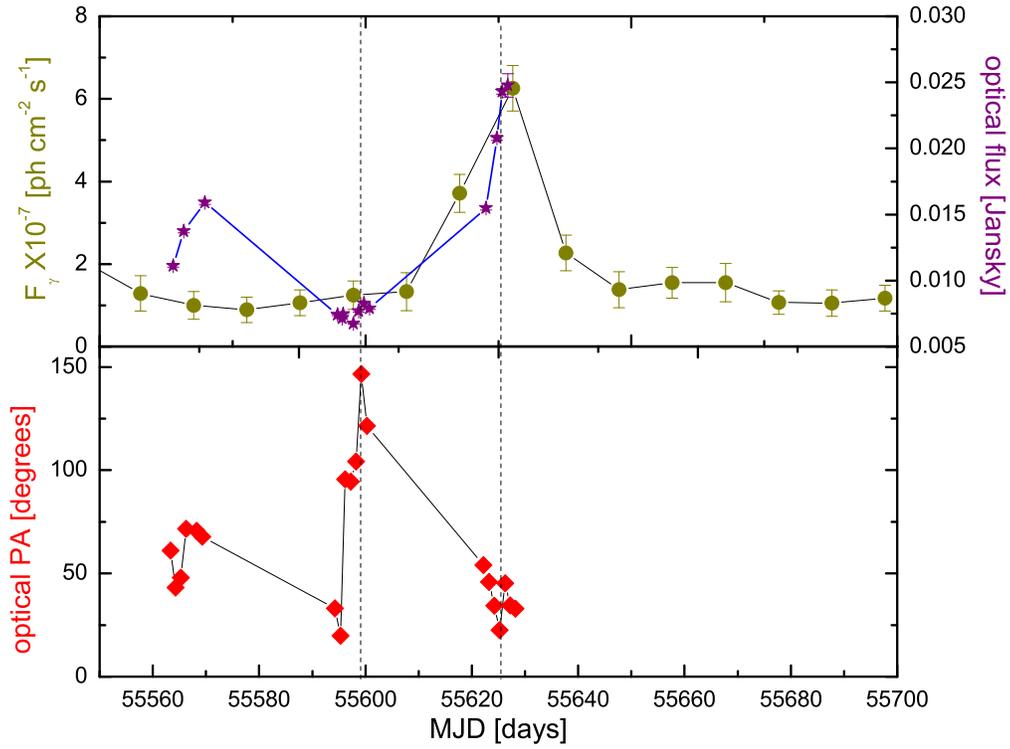}
 \caption{The coincidence of $\gamma$-ray and optical flares with a dramatic change of V band PA. The dark yellow filled circles represent the $\gamma$-ray fluxes, the purple stars are the optical fluxes and the red squares represent the V band PA. Between the two vertical dash lines, the fluxes of the $\gamma$-ray and the optical increase while the PA shows the overall downward trend.}
  \label{Figure 5.}
\end{figure}

\begin{figure*}
\centering
\includegraphics[width=4.5 in,angle=-90]{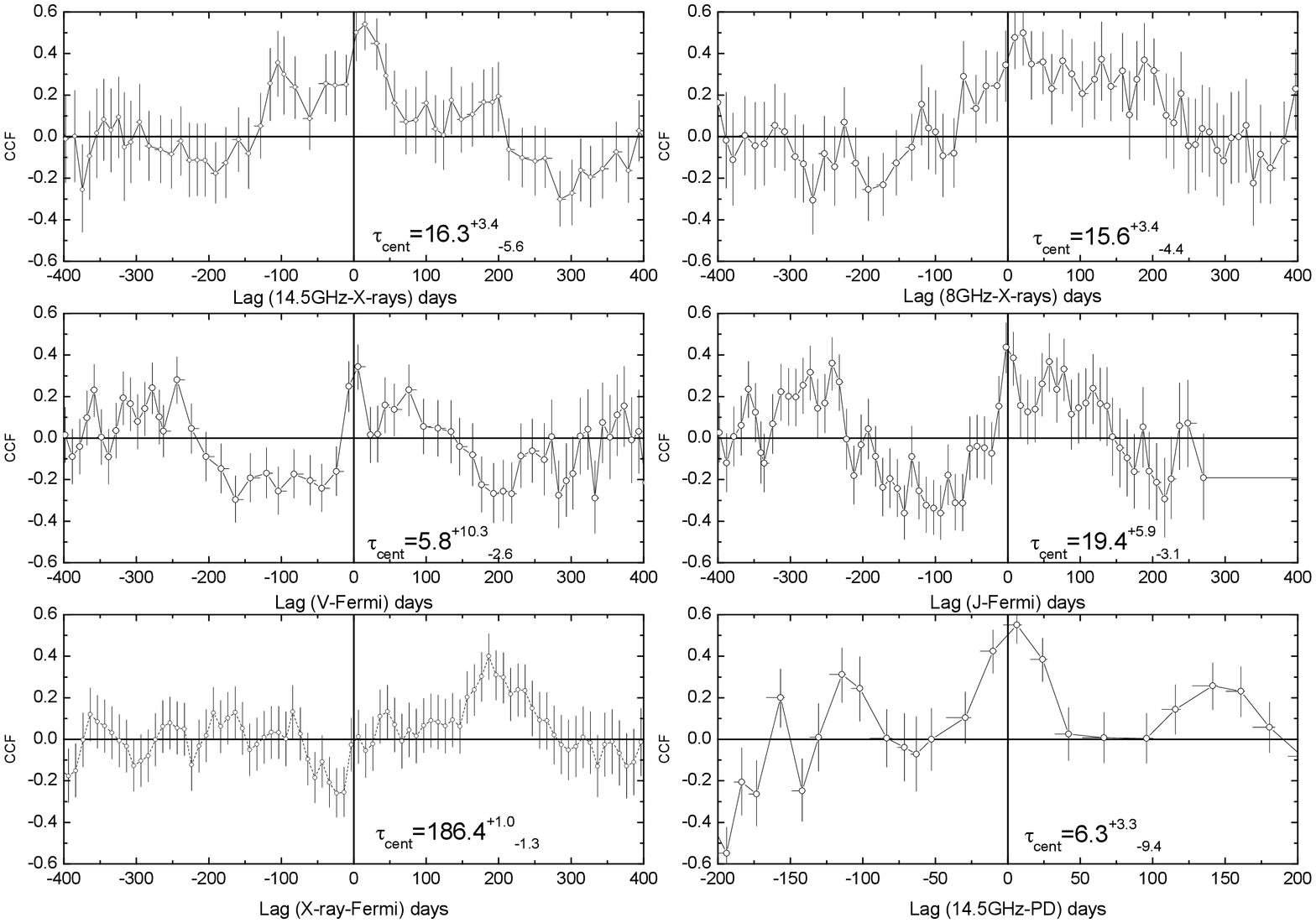}
 \caption{ZDCFs between $\gamma$-rays, X-rays, V band, J band, 14.5 GHz and 8 GHz emissions, together with V band PD. Considering the bin sizes of the PCA X-ray and LAT $\gamma$-ray light curves, radio/hard X-ray and the optical/$\gamma$-ray variations are suggested to be correlated with zero-lag}
  \label{Figure 6.}
\end{figure*}

\begin{figure}
\centering
\includegraphics[width=3.5 in,angle=-90]{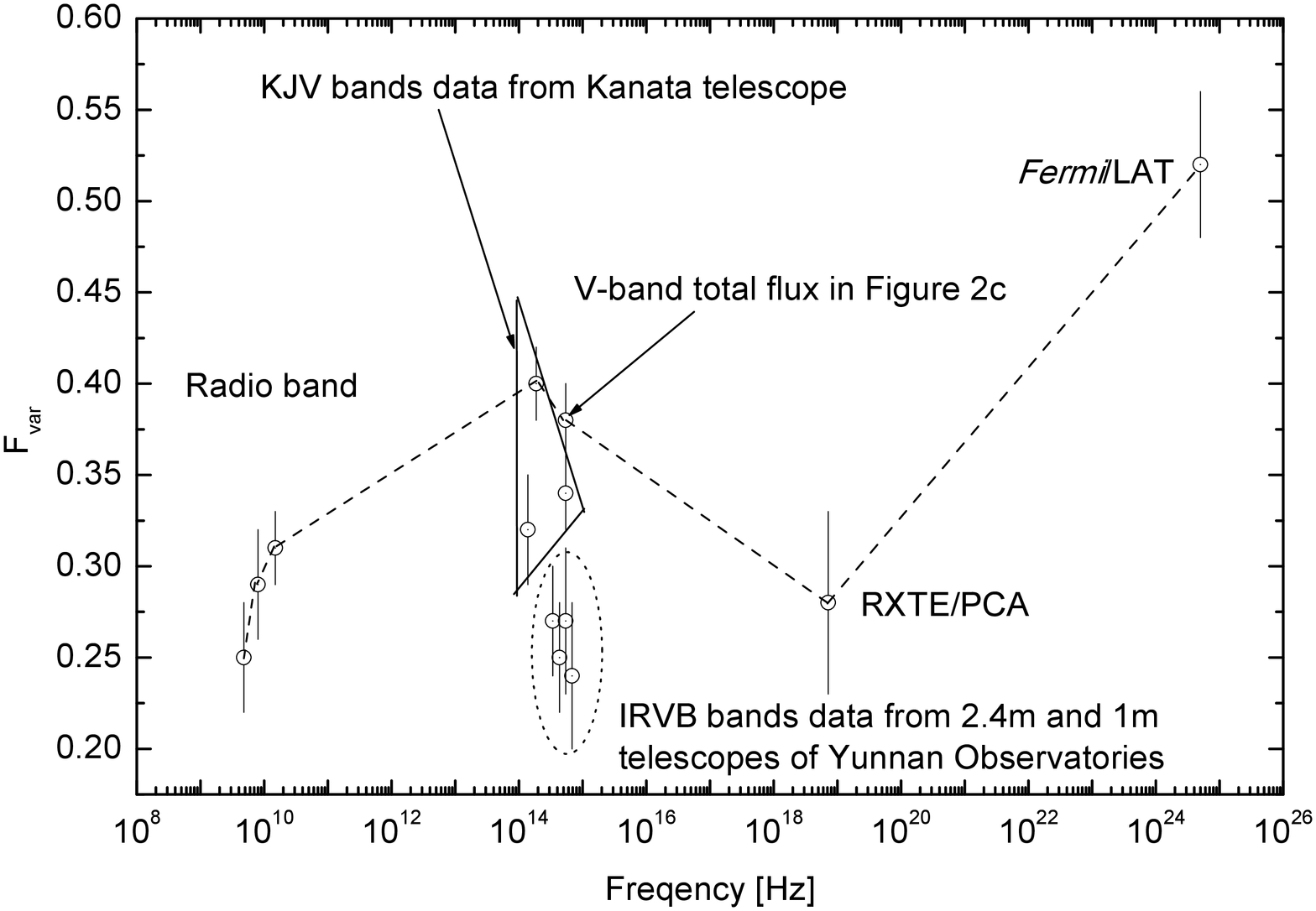}
 \caption{Fractional variability amplitude as function of frequency. The dashed line shows
 the data in Table 2.}
  \label{Figure 7.}
\end{figure}

\begin{figure}
\centering
\includegraphics[width=3.5 in,angle=0]{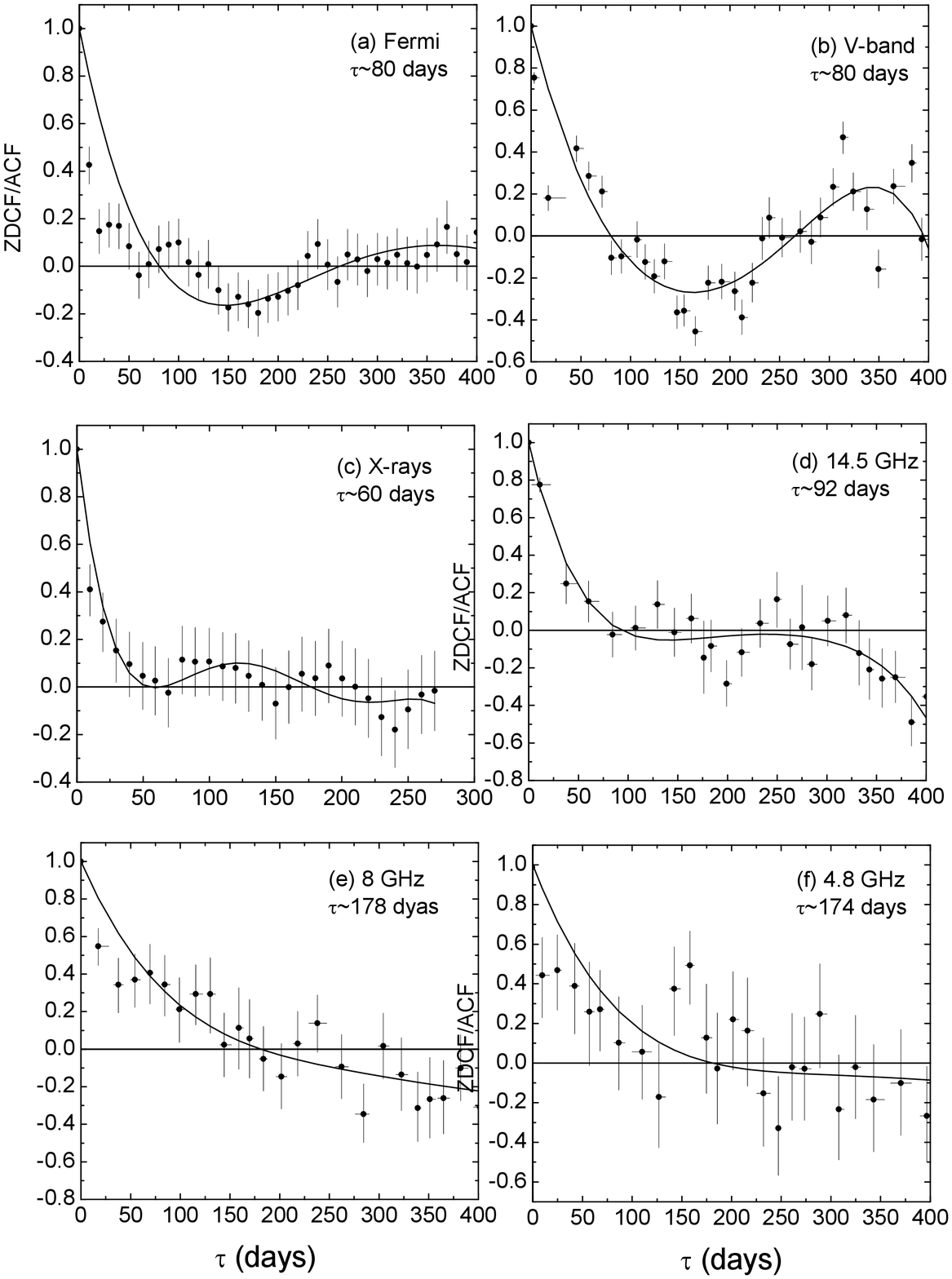}
 \caption{The ZDCFs (discrete points) and their fits (fifth-order polynomial
least-squares fit, solid line).}
  \label{Figure 8.}
\end{figure}

\begin{figure}
\centering
\includegraphics[width=5.5 in]{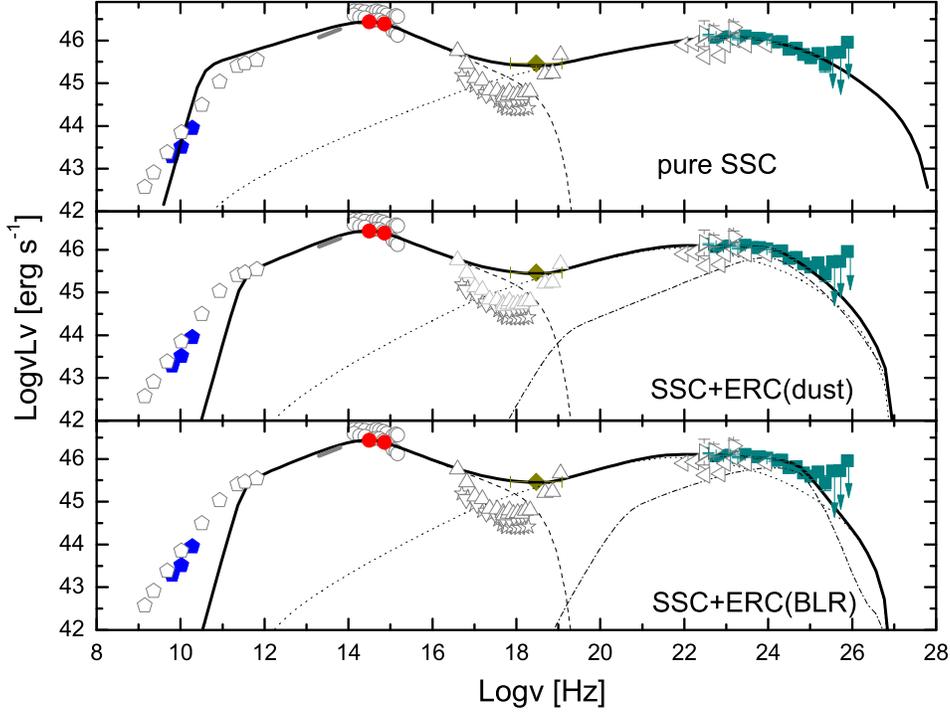}
 \caption{Modeling the average SED with different models. Filled symbols are from our work and hollow symbols are collected from the literatures (radio, Fuhrmann et al. 2008; IR/optical, Chen $\&$ Shan 2011, Villata et al 2008; X-ray, Tagliaferri et al. 2003; $\gamma$-ray, Tagliaferri et al. 2003, Giommi et al. 2008). Blue pentagons are the averaged fluxes of 4.8, 8, 14.5 GHz from UMRAO. Red filled circles are the J and V band optical average fluxes. Yellow diamond is the average X-ray flux for 3-50 keV from RXTE/PCA. Cyan squares represent the 40 months average $\gamma$-ray spectrum from {\it Fermi}/LAT. The dash line represents the calculated synchrotron emission, the dot line corresponds to the SSC component and the dash dot line is the ERC part.}
  \label{Figure 9.}
\end{figure}

\begin{figure}
\centering
\includegraphics[width=3.5 in,angle=0]{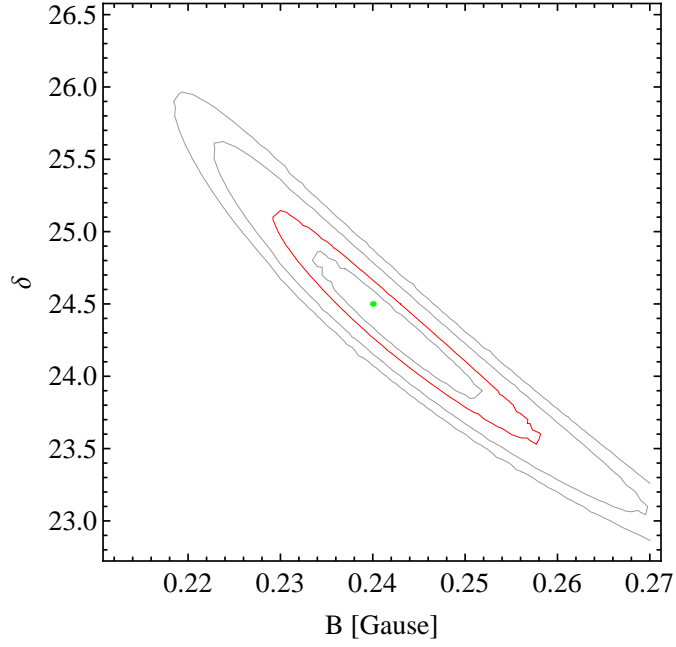}
 \caption{The contours of the p in the B-$\delta$ plane, while $p\propto e^{-\chi^{2}/2}$. Four contours correspond to the p of 0.05, 0.1, 0.2 and 0.26, respectively. And the contour of p=0.2 is drawn as red color, which corresponds to the 1$\sigma$ level of the Gaussian fit for the B-p and $\delta$-P distribution. The green spot demonstrates the location of the best fit.}
 \label{Figure 10.}
\end{figure}

\begin{figure}
\centering
\includegraphics[width=3.2 in]{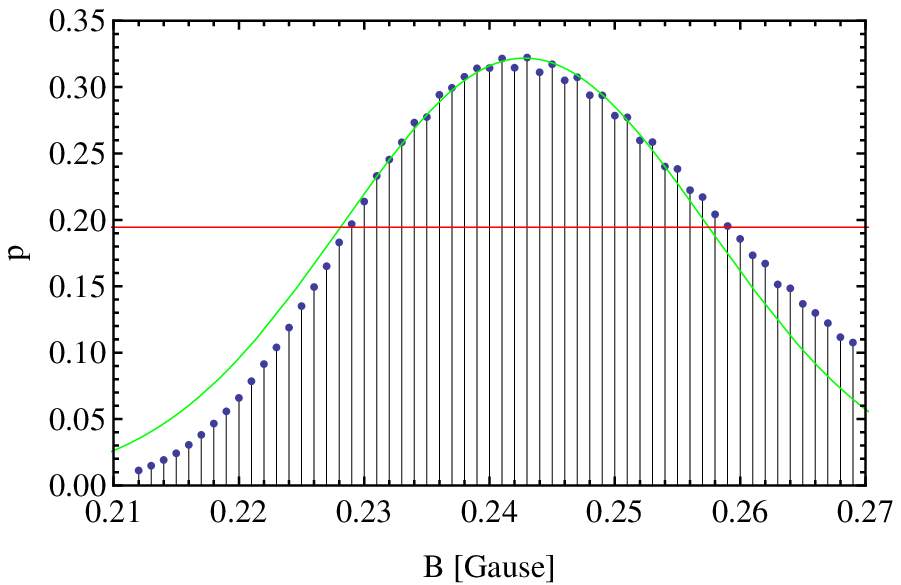}
\includegraphics[width=3.2 in]{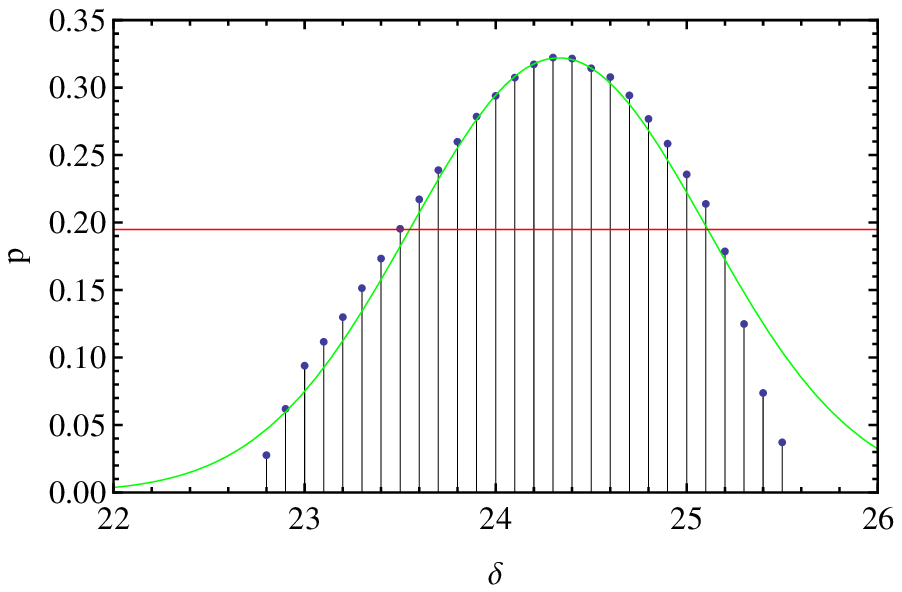}
 \caption{The distributions of p with different values of B and $\delta$. The green lines are the Gaussian fits. The red horizontal lines are obtained from the 1$\sigma$ level of the Gaussian fits with the p value of 0.195. The 1$\sigma$ uncertainties of the B and $\delta$ are obtained from the intersections between the B-p and $\delta$-p distributions and the red horizontal lines.}
  \label{Figure 11.}
\end{figure}

\begin{figure}
\centering
\includegraphics[width=5.5 in]{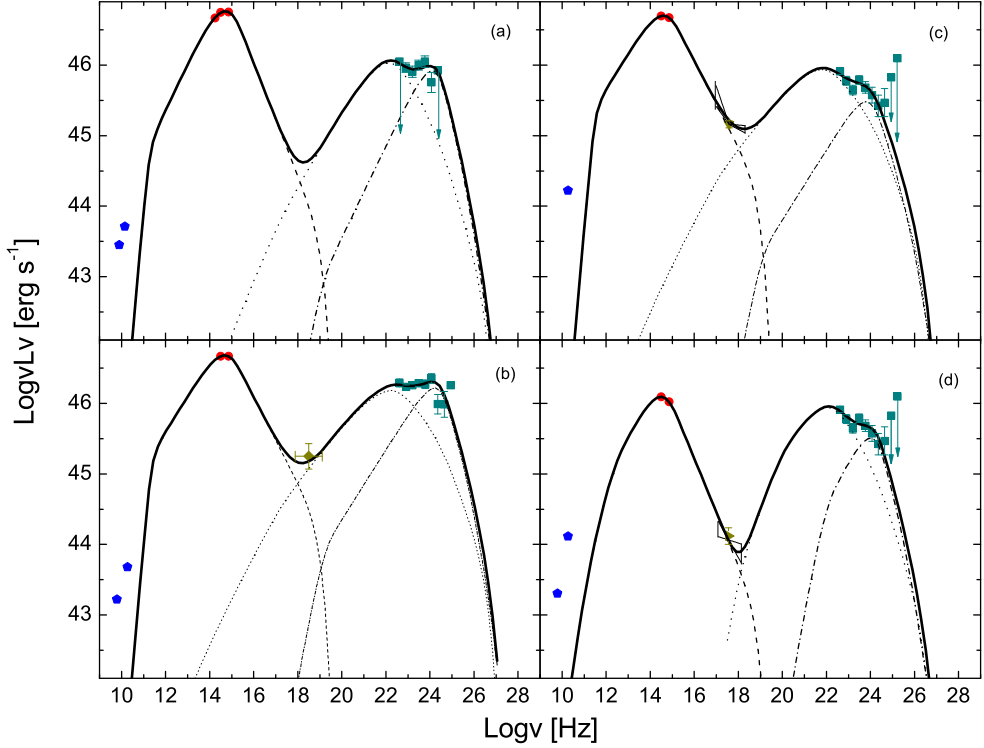}
 \caption{Modeling the four simultaneous SEDs. Blue pentagons are the radio observations from UMRAO. Red filled circles are the optical and NIR photometric data. The error bars for radio and optical fluxes are too small to be seen. Yellow diamond in (b) is the interpolation by nearby RXTE/PCA observations. The X-ray spectra with yellow triangles is from {\it Swift}/XRT. The cyan squares are the {\it Fermi}/LAT observations. The dash lines represent the calculated synchrotron emissions, the dot lines correspond to the SSC components and the dash dot lines are the ERC parts. (a) is the SED at 2008 Dec. with the strongest optical flare and a $\gamma$-ray flare with medium peak flux. (b) represents the SED at 2009 Oct. when both optical and $\gamma$-ray peak fluxes are in high state. (c) corresponds with a multi-wavelength flaring state at 2009 Dec. but the peak flux of $\gamma$-ray flare is extreme low. (d) is the SED that all bands are in low state at 2010 Jan.}
  \label{Figure 12.}
\end{figure}

\begin{figure}
\centering
\includegraphics[width=5.5 in,angle=0]{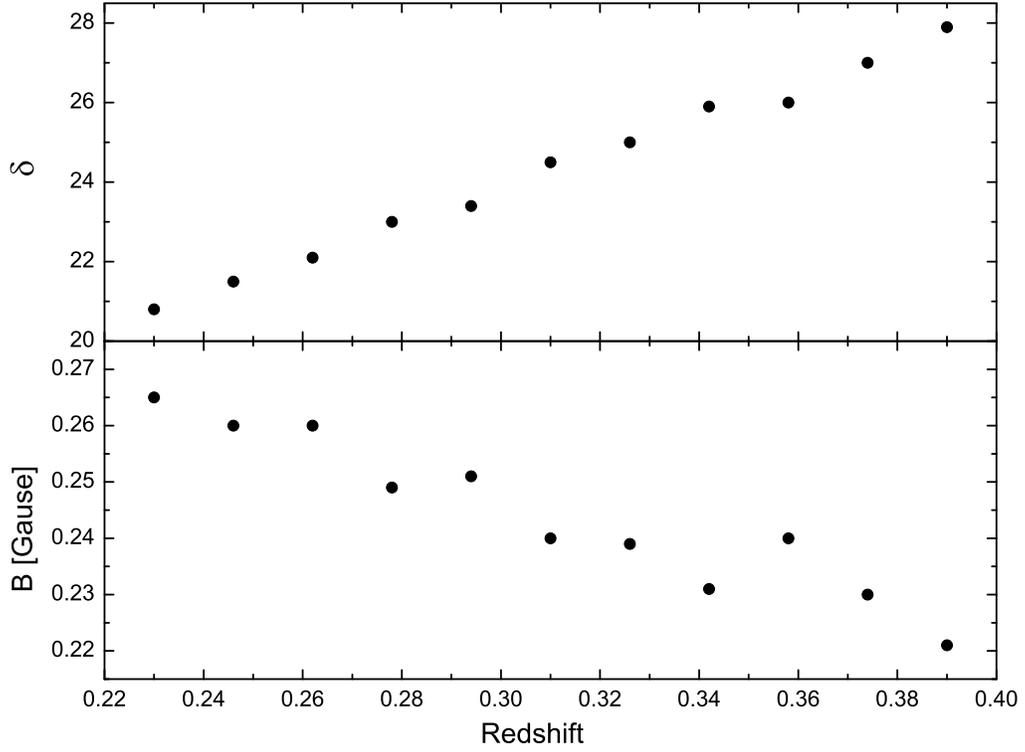}
 \caption{The distributions of the B and $\delta$ with different values of redshift. 11 redshift points are chosen evenly in the 1$\sigma$ uncertainty of S5 0716+714 from Nilsson et al. (2008). The values of the B and $\delta$ are obtained by modeling 11 SEDs with different redshift independently using the SSC+ERC model with hot dust emission as the external photons.}
  \label{Figure 13.}
\end{figure}

\clearpage

\begin{deluxetable}{lcccccccc}
\tablenum{1}\tablecolumns{6} \tabletypesize{\scriptsize} \tabcolsep 5pt
\tablewidth{0pt} \tablecaption{Results of {\it Swift} spectral fits
during different state\tablenotemark{a}}

\tablehead{ \colhead{ } &\colhead{ } &\colhead{} &\colhead{}
&\colhead{ }
&\colhead{Flux\tablenotemark{b}}  \\

\colhead{Model} &\colhead{$\Gamma_1$} &\colhead{$E_{\rm
break}$(keV)} &\colhead{$\Gamma_2$} &\colhead{$\chi^2_{\nu}$/dof}
&\colhead{(0.3--10~keV)} }

\startdata


\noalign{\smallskip} \hline \noalign{\smallskip}
&& 2009-11-30 to 2009-12-02 (outburst)&&&\\
\noalign{\smallskip} \hline \noalign{\smallskip}

Single power law &$2.30^{+0.14}_{-0.14}$ &--  &--  &1.27/33  &2.71 \\


\noalign{\smallskip} \hline \noalign{\smallskip}
&& 2009-12-07 to 2009-12-22 (falling)&&&\\
\noalign{\smallskip} \hline \noalign{\smallskip}

Single power law &$2.21^{+0.03}_{-0.03}$ &--  &--  &1.36/102  &1.60 \\

Broken power law & $2.32^{+0.06}_{-0.06}$  &$1.27^{+0.34}_{-0.23}$ &$2.09^{+0.06}_{-0.07}$ &1.26/100 & 1.66 \\

Double power law &$2.32^{+0.71}_{-0.10}$ &-- &$0.92^{+2.11}_{-0.92}$ &1.29/100  &1.69 \\


\noalign{\smallskip} \hline \noalign{\smallskip}
&& 2009-12-28 to 2010-02-22 (quiet)&&&\\
\noalign{\smallskip} \hline \noalign{\smallskip}

Single power law &$2.33^{+0.24}_{-0.23}$ &--  &--  &1.39/21  &0.12 \\


\noalign{\smallskip} \hline \noalign{\smallskip}
&& 2011-07-14 to 2011-07-16 &&\\
\noalign{\smallskip} \hline \noalign{\smallskip}

Single power law &$2.09^{+0.15}_{-0.14}$ &-- &--  &1.42/69  &1.66 \\

Broken power law & $2.49^{+0.51}_{-0.23}$  &$3.26^{+0.52}_{-1.07}$ &$1.17^{+0.56}_{-0.44}$ &1.20/67 & 2.30 \\

Double power law &$0.71^{+0.84}_{-1.79}$ &-- &$3.31^{+1.73}_{-0.84}$ &1.18/67  &3.76 \\

\noalign{\smallskip} \hline \noalign{\smallskip}
&& 2011-10-25 to 2011-10-28 &&&\\
\noalign{\smallskip} \hline \noalign{\smallskip}

Single power law &$2.82^{+0.26}_{-0.23}$ &--  &--  &1.96/46  &1.33 \\

Broken power law & $3.88^{+0.66}_{-0.55}$ &$2.21^{+0.58}_{-0.35}$ &$1.85^{+0.43}_{-0.58}$ &1.51/44 & 4.04 \\

Double power law &$1.40^{+0.71}_{-1.05}$ &-- &$5.06^{+1.84}_{-1.27}$ &1.65/44  &1.36 \\

\enddata
\tablenotetext{a}{The fits are performed in the 0.3--10~keV band.
The neutral hydrogen absorption column density is fixed to the Galactic value for the data during 2009. However, the column density is treated as a free parameter for 2011 data.
All quoted errors are $90\%$ confidence level
($\Delta\chi^2=2.706$) for one interesting parameter. Two X-ray spectra at 2009 are used for modeling the simultaneous SEDs . Other three X-ray spectra are not used due to lacking of simultaneous optical data.}

\tablenotetext{b}{The unabsorbed flux is in unit of
$\rm{10^{-11}~erg ~ cm^{-2}~ s^{-1}}$.}

\label{tab:spec}
\end{deluxetable}

\begin{deluxetable}{lrrrrrr}
\tablenum{2} \tablewidth{0pt} \tabletypesize{\scriptsize}
\tabcolsep 5pt \tablecaption{The fractional
variability amplitudes $F_{\rm{var}}$ for radio, infrared, optical,
X-ray, and $\gamma$-ray light curves}

\tablehead{ \colhead{4.8 GHz} & \colhead{8GHz}&\colhead{15 GHz} &\colhead{J-band} &\colhead{V-band}&\colhead{X-ray}&\colhead{{\it Fermi}}\\
\colhead{(1)}&\colhead{(2)}&\colhead{(3)}&\colhead{(4)}&\colhead{(5)}&\colhead{(6)}&\colhead{(7)}}

\startdata

$0.25\pm 0.03$&$0.29\pm 0.03$&$0.31\pm 0.02$&$0.40\pm 0.02$&$0.38\pm 0.02$&$0.28\pm 0.05$&$0.52\pm 0.04$\\

\enddata

\end{deluxetable}

\begin{deluxetable}{lcccccccccccc}
\tablenum{3}\tabcolsep 3pt
\tabletypesize{\scriptsize}
\tablewidth{0pt} \tablecaption{Input Parameters of the SED Model\tablenotemark{a}}
\tablehead{\colhead{Parameters\tablenotemark{1}} &\colhead{Pure SSC} &\colhead{SSC+ERC(BLR)} &\colhead{SSC+ERC(dust)} &\colhead{Dec.2008} &\colhead{Oct.2009} &\colhead{Dec. 2009} &\colhead{Jan. 2010\tablenotemark{b}}}

\startdata

B(Gauss)&$6_{+1}^{-2}\times10^{-3}$ &$0.28_{+0.02}^{-0.02}$ &$0.24_{+0.02}^{-0.01}$ &$0.34_{+0.06}^{-0.05}$ &$0.22_{+0.05}^{-0.03}$ &$0.48_{+0.1}^{-0.07}$ &$0.19_{+0.03}^{-0.03}$\\[3pt]
$\delta$ &$39.6_{+3.8}^{-3.1}$ &$23.3_{+0.7}^{-0.9}$ &$24.5_{+0.6}^{-0.7}$ &$27.0_{+4.8}^{-3.1}$ &$28.0_{+2.7}^{-3.0}$ &$24.1_{+2.3}^{-1.7}$ &$20.3_{+2.0}^{-1.4}$ \\[3pt]
$t_{var}$(day) &10 &1 &1 &1 &1 &1 &1\\[3pt]
R(cm) &$7.8\times10^{17}$ &$4.6\times10^{16}$ &$4.8\times10^{16}$ &$5.3\times10^{16}$ &$5.5\times10^{16}$ &$4.8\times10^{16}$ &$4.0\times10^{16}$\\[4pt]
K &$1.2\times10^{3}$  &$1.1\times10^{4}$ &$1.0\times10^{4}$ &$0.1\times10^{3}$ &$1.4\times10^{3}$ &$1.3\times10^{3}$ &$0.9\times10^{3}$ \\[3pt]
$\gamma_{br}$  &$2.0\times10^{4}$  &$3.3\times10^{3}$ &$3.5\times10^{3}$ &$4.0\times10^{3}$ &$4.7\times10^{3}$ &$3.1\times10^{3}$ &$4.8\times10^{3}$\\[3pt]
$p_{1}$  &2.4  &2.2 &2.2 &1.7 &2.0 &2.0 &1.8\\[3pt]
$p_{2}$  &3.9  &3.8 &3.8 &4.6 &4.2 &4.2 &4.5\\[3pt]

\enddata

\tablenotetext{a}{Seven SEDs have been modeled in our work and the input parameters are listed in this table. Three groups of parameters are obtained from modeling the average SEDs by different models. Other four groups of parameters with dates are obtained from the modeling simultaneous SEDs with synchrotron plus SSC+EC(IR) model containing three flaring state and one low state at 2010 Jan.}
\tablenotetext{b}{Detailed description of the multi-wavelength variability corresponding to the four simultaneous SEDs are shown in section 5.2 in this paper.}
\tablenotetext{1}{ Input parameters are described in the first paragraph of section 5. For all fits except the case of Jan. 2010, the $\gamma_{\rm min}$ is set to 10. For the fit of Jan. 2010 the $\gamma_{\rm min}$ is chosen as 350 due to its extremely low X-ray flux. $\gamma_{\rm max}$ are 100 times as the $\gamma_{\rm min}$ for all cases.}

\end{deluxetable}

\clearpage

\appendix
\section{The multi-bands photometric data from Yunnan Observatories}
\begin{deluxetable}{lrrrrrr}
\tablenum{4} \tablewidth{0pt}
\tablecaption{The multi-bands photometric data from Yunnan Observatories\tablenotemark{1}}
\tablehead{ \colhead{MJD\tablenotemark{2}} &\colhead{Mag.\tablenotemark{3}} &\colhead{SigMag.\tablenotemark{4}} &\colhead{Band\tablenotemark{5}} }

\startdata
54731.9 &12.674  &0.009 &I  \\[3pt]
54733.9 &12.563  &0.016 &I  \\[3pt]
\enddata
\tablenotetext{1}{Table 4 is available in its entirety in machine-readable forms in the online journal.
A portion is shown here for guidance regarding its form and content.}
\tablenotetext{2}{The observation date}
\tablenotetext{3}{The nightly average magnitude. The correction for the interstellar extinction has been already completed}
\tablenotetext{4}{Uncertainty of magnitude}
\tablenotetext{5}{The photometric band }
\end{deluxetable}

\end{document}